\documentclass{osa-article}
\journal{oe}

\usepackage{graphicx}
\usepackage{bm}
\usepackage{color}
\usepackage{amsmath}

\DeclareMathOperator{\sech}{\textrm{sech}}
\DeclareMathOperator{\sign}{\textrm{sign}}
\DeclareMathOperator{\const}{\textrm{const}}

\articletype{Research Article}

\begin{document}

\title{Bose-Einstein condensate soliton qubit states for metrological applications}

\author{The Vinh Ngo,\authormark{1} Dmitriy Tsarev,\authormark{1} Ray-Kuang Lee,\authormark{2,3,4} and Alexander Alodjants\authormark{1,*}}

\address{%\authormark{1}National Research University for Information Technology, Mechanics and Optics (ITMO), St. Petersburg 197101, Russia\\
\authormark{1}Faculty of Laser Photonics and Optoelectronics, ITMO University, St. Petersburg 197101, Russia\\
\authormark{2}Physics Division, National Center for Theoretical Sciences, Hsinchu 30013, Taiwan\\
\authormark{3}Institute of Photonics Technologies, National Tsing Hua University, Hsinchu 30013, Taiwan\\
\authormark{4}Center for Center for Quantum Technology, Hsinchu 30013, Taiwan}

\email{\authormark{*}alexander\_ap@list.ru} %% email address is required

\begin{abstract}
By utilizing Bose-Einstein condensate solitons, optically manipulated and trapped in a double-well potential, coupled through nonlinear Josephson effect, we propose novel quantum metrology applications with two soliton qubit states. In addition to steady-state solutions in different scenarios, phase space analysis, in terms of population imbalance - phase difference variables, is also performed to demonstrate macroscopic quantum self-trapping regimes. Schr\"odinger-cat states, maximally path-entangled ($N00N$) states, and macroscopic soliton qubits are predicted and exploited for the distinguishability of obtained macroscopic states in the framework of binary (non-orthogonal) state discrimination problem. For arbitrary phase estimation in the framework of linear quantum metrology approach, these macroscopic soliton states are revealed to have a scaling up to the Heisenberg limit (HL). The examples are illustrated for HL estimation of angular frequency between the ground and first excited macroscopic states of the condensate, which opens new perspectives for current frequency standards technologies.
\end{abstract}

\section{Introduction}
Nowadays, nonlinear collective mode formation and interaction in Kerr-like medium represent an indispensable platform for various practically important applications in time and frequency metrology~\cite{Hansch, Papp}, spectroscopy~\cite{Suh, Dutt}, absolute frequency synthesis~\cite{Spencer}, distance ranging~\cite{Trocha}. In photonic settings frequency combs are proposed for these purposes~\cite{Gaeta}. The combs occur due to the nonlinear mode mixing in special (ring) microcavities, which possess some certain eigenmodes. Notably, bright soliton formation emerges with vital phenomena accompanying micro-comb generation~\cite{Kippenberg1}. Physically, such a soliton arises due to the purely nonlinear effect of temporal self-organization pattern occurring in an open (driven-dissipative) photonic system. However, because of the high level of various noises in the system they can be hardly explored for purely quantum metrological purposes. 

Instead, atomic optics, which operates with Bose-Einstein condensates (BECs) at low temperatures, provides a suitable platform for various quantum devices that may be useful for metrology and sensing tasks~\cite{Pezze}. In particular, so-called Bosonic Josephson junction (BJJ) systems, established through two weakly linked and trapped atomic condensates, are at the heart of the current quantum technologies in atomtronics that considers atom condensates and aims to design (on-chip) quantum devices. The condensates in this case represent low dimensional systems and may be manipulated by magnetic and laser field combinations. In this sense, they represent an alternative to optical analogues. 
 
The BJJs are intensively discussed and examined both in theory and experiment~\cite{Anthony2001,Raghavan1999,Ananikian2006,Sols2002,Gati2007}. The quantum properties of the BJJs are also widely studied~\cite{Cirac1998,Reid2011, Reid2012, Olsen2010, Salasnich2011, Reid2018, Sorensen2001, Sorensen2001a}, demonstrating spin-squeezing and entanglement phenomena~\cite{He, Puentes, Vitagliano2018}, with the capability in generating $N00N$-states~\cite{Mazzarella, Olsen} to go beyond the standard quantum limit~\cite{Toth}. Physically, the BJJs possess interesting features connected with the interplay between quantum tunneling of the atoms and their nonlinear properties evoked by atom-atom interaction~\cite{Ostrovskaya2000,Elyutin2001}. 

With the Kerr-like nonlinearities, solitons naturally emerge from atomic condensates in low dimensions, see e.g.~\cite{Kevrekidis2008, Eiermann2004,Strecker2002, Khaykovich2002, Nguyen2014}.
%Roughly speaking in this limit, inter-particle interaction may be strong enough and capable for manipulation of atomic scattering length by means of magnetic field. 
Especially, the bright atomic solitons observed in lithium condensate possessing negative scattering length~\cite{Strecker2002, Khaykovich2002, Nguyen2014} are worth noticing. Atomic gap solitons are also observed in condensates with repulsive inter-particle interaction~\cite{Eiermann2004}. Based on soliton modes, we recently proposed the quantum soliton Josephson junction (SJJ) device with the novel concept to improve the quantum properties of effectively coupled two-mode system~\cite{Tsarev2018,Tsarev2019,Tsarev2020}. The SJJ-device consists of two weakly-coupled condensates trapped in a double-well potential and elongated in one dimension, cf.~\cite{Raghavan2000}. We demonstrated that quantum solitons may be explored for the improvement of phase measurement and estimation up to the Heisenberg limit (HL) and beyond~\cite{Tsarev2018}. In the framework of nonlinear quantum metrology, we also showed that solitons permits Super-Heisenberg (SH) scaling ($\propto N^{-3/2}$) even with coherent probes~\cite{Tsarev2019}. On the other hand, steady-states of coupled solitons can be useful for effective formation of Schr{\"o}dinger-cat (SC) superposition state and maximally path-entangled $N00N$-states, which can be applied for the phase estimation purposes\cite{Tsarev2020}. It is important that such superposition states arise only for soliton-shape condensate wave functions and occur due to the existence of certain steady-states in phase difference - population imbalance phase plane, cf.~\cite{Tsarev2019}. 

Remarkably, macroscopic states, like SC-states, play an essential role for current information and metrology, see e.g.~\cite{Gilchrist2004}. In quantum optics, various strategies are proposed for creation photonic SC-states and relevant (continuous variable) macroscopic qubits~\cite{Ourjoumtsev2006, Neergaard2010, Izumi2018}. Special (projective) measurement and detection techniques are also important here~\cite{Lundeen2009, Takeoka2005, Lvovsky2020}. From this point of view, the fact that condensate environment, dealing with mater waves, provides the minimally accessible thermal noises makes it potentially promising for macroscopic qubits implementation, cf.~\cite{Byrnes2012,Demirchyan2014,Sinatra2017}.
%\textcolor{red}{condensate environment dealing}
%with matter waves provides the minimally accessible thermal noises, which is potentially promising for macroscopic qubits implementation, cf.~\cite{Byrnes2012,Demirchyan2014,Sinatra2017}.

In this work, we propose metrological applications for two soliton superposition states as macroscopic qubits. The interaction between these solitons comes from the nonlinear mode mixing in an atomic condensate trapped in a double-well potential. In particular, we reveal the SC-states formation and their implementation for arbitrary phase measurement prior HL and beyond. Since SC-states are non-orthogonal states, a special measurement procedure is applied by so-called sigma operators, as it enables us to estimate the unknown phase parameter~\cite{Gerry2007}. On the other hand, our approach can be also useful in the framework of discrimination of binary coherent (non-orthogonal) states in quantum information and communication, see e.g.~\cite{Sasaki1996,Loock2005}. The non-orthogonality of these states leads to so-called Helstrom bound for the quantum error probability that simply indicates the impossibility for a receiver to identify the transmitted state without some errors~\cite{Helstrom1976,Geremia2004}. In quantum metrology, by means of various regimes of condensate soliton interaction, we deal with a set of states for a quantum system, which may be prepared before the measurement. Our results show that these SC-states approach the soliton $N00N$-states to minimize the quantum error probability. 

\section{Two-soliton model}

\subsection{Coupled-mode theory approach}

We start with the mean-field description of coupled mode-theory approach to elongated (in $X$-dimension) BEC trapped in $V=V_H+V(x)$ potential, where $V_H$ is a familiar 3D harmonic trapping potential, $V(x)$ is responsible for condensate double-well confinement in one ($X$) dimension~\cite{Tsarev2020}. The condensate (rescaled) wave function (mean field amplitude) $\Psi(x)$ obeys the familiar 1D Gross-Pitaevskii Equation (GPE), cf.~\cite{Raghavan2000}:
\begin{equation}\label{GP00}
i\frac{\partial}{\partial t}\Psi = -\frac{1}{2}\frac{\partial^2}{\partial x^2}\Psi - u\left|\Psi\right|^2\Psi + V(x)\Psi,
\end{equation}
where $u=4\pi|a_{sc}|/a_{\perp}$ characterizes Kerr-like (focusing) nonlinearity, $a_{sc}<0$ is the s-wave scattering length that appears due to atom-atom scattering in Born-approximation, $a_{\perp}=\sqrt{\hbar/m\omega_{\perp}}$ characterizes the trap scale, and $m$ is the particle mass. To be more specific, we only consider condensates possessing a negative scattering length. In Eqs.~\eqref{GP00} we also propose rescaled (dimension-less) spatial and time variables, which are $x,y,z \rightarrow x/a_{\perp},y/a_{\perp},z/a_{\perp}$, and $t \rightarrow \omega_{\perp} t$, cf.~\cite{Tsarev2018,Tsarev2019,Raghavan2000}. 

The nonlinear coupled-mode theory admits solution of~\eqref{GP00} that simply represents a quantum-mechanical superposition 
\begin{equation}\label{sup}
\Psi(x,t)=\Psi_1(x,t)+\Psi_2(x,t),
%i\frac{\partial}{\partial t}\Psi = -\frac{1}{2}\frac{\partial^2}{\partial x^2}\Psi - U\left|\Psi\right|^2\Psi,
\end{equation}
where wave functions $\Psi_{1}(x)$ and $\Psi_{2}(x)$ characterize the condensate in two wells. For weakly interacting atoms one can assume that 
\begin{equation}\label{wf1}
\Psi_{1,2}(x,t)=C_{1,2}(t)\Phi_{1,2}(x)e^{i\beta_{1,2}t},
%i\frac{\partial}{\partial t}\Psi = -\frac{1}{2}\frac{\partial^2}{\partial x^2}\Psi - U\left|\Psi\right|^2\Psi,
\end{equation}
where $\Phi_1(x)$ and $\Phi_2(x)$ are ground- and first-order excited mode state wave functions possessing energies $\beta_1$ and $\beta_2$, respectively; $C_{1}(t)$ and $C_{2}(t)$ are time-dependent functions. If the particle number is not too large, GPE~\eqref{GP00} may be integrated in spatial ($X$ - dimension) leaving only two condensate variables $C_{1,2}(t)$, cf.~\cite{Elyutin2001}. In particular, $\Phi_1(x)$ and $\Phi_2(x)$ may be time-independent Gaussian-shape wave functions obeying different symmetry. Practically, this two-mode approximation is valid for the condensates of several hundreds of particles~\cite{Anglin2001}. The condensate in this limit is effectively described by two (macroscopically populated) modes as a result. 

%The phase portrait for population imbalance and phase difference exhibits various dynamical regimes for coupled condensates. However, how we will see below, ..... 

\subsection{Quantization of coupled solitons}

\begin{figure}[h!]
\begin{center}
\includegraphics[width=0.5\linewidth]{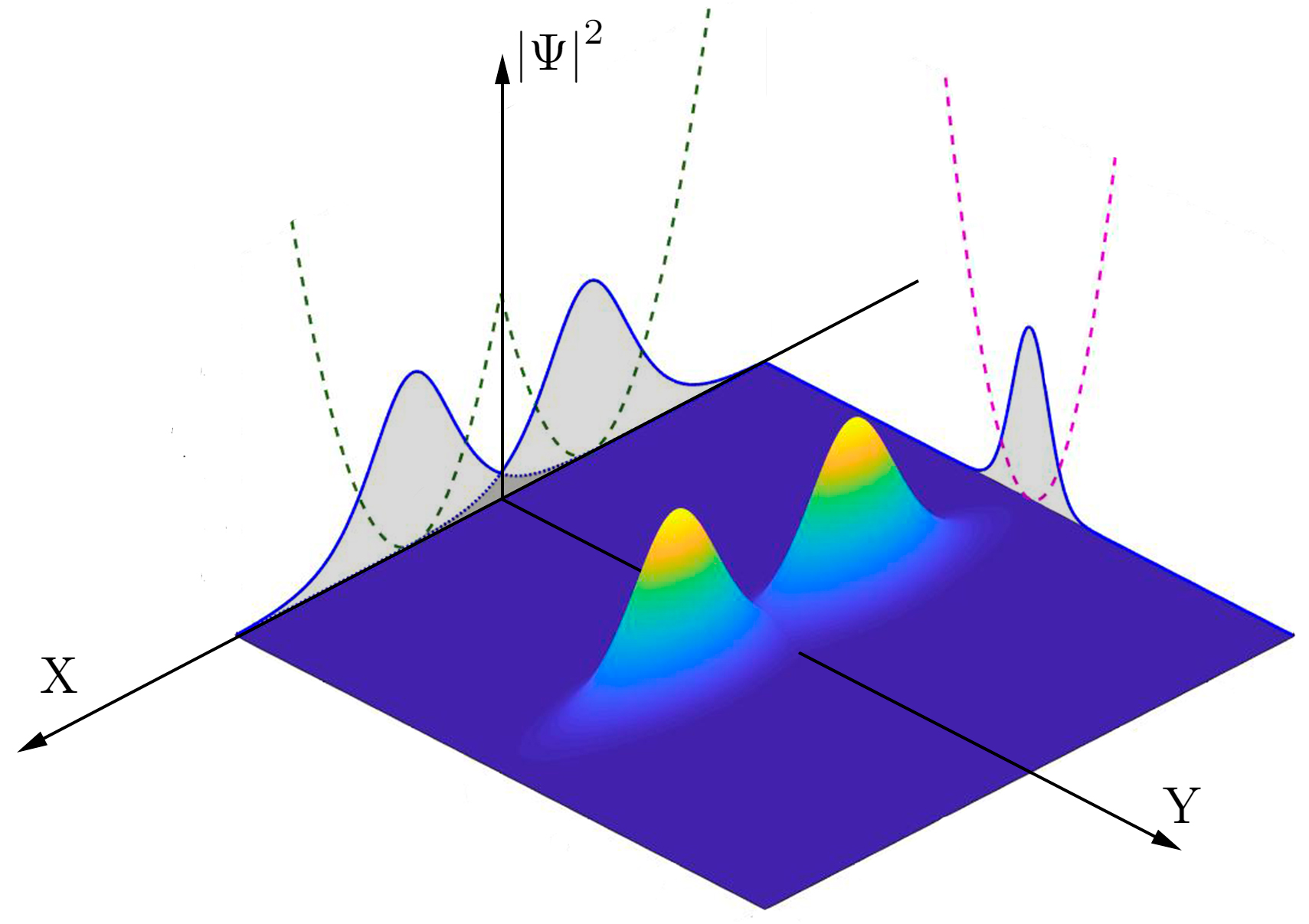}
\end{center}
%\caption{Sketch of probability density distribution $|\Psi|^2$ versus spatial coordinates $X$ and $Y$ for the coupled condensates trapped in a double-well (dashed green curve) and harmonic (dashed magenda curve) potentials, respectively. Shadow regions display condensate wave packets projections; they represent secant-shape in $X0Z$-plane, and Gaussian-shape in $Y0Z$ one, respectively.}
\caption{Sketch of probability density distribution $|\Psi|^2$ versus spatial coordinates $X$ and $Y$, 2D projection of the 3D coupled condensates trapped in a double-well (dashed green curve) and harmonic (dashed magenda curve) potentials, respectively. Shadow regions display 1D condensate wave packets projections; they represent a secant-shape in $X$-direction, and Gaussian-shape in the transverse directions.}
\label{FIG:Solitons}
\end{figure}

The sketch in Fig.~\ref{FIG:Solitons} explains the two-soliton system described in our work. If trapping potential $V(x)$ is weak enough and the condensate particles interact not so weakly, the ansatz solution~\eqref{wf1} is no more suitable. Especially we would like to mention condensates with a negative scattering length which admit a bright soliton solution for $\Psi_{1,2}(x,t)$ in~\eqref{sup}. In fact, in this case one can speak about two-soliton solution problem for GPE~\eqref{GP00} without trapping potential $V(x)$ known in classical theory of solitons~\cite{Karpman1981}. 

%Physically, two-soliton superposition state $\Psi(x)$ can be understood in the framework of nonlinear collective mode theory and corresponds to the ground and first-order excited state superposition in BEC,~\cite{4}.

In quantum theory we deal with a bosonic field operator $\hat{a}(x,t)\propto \hat{a}_1+\hat{a}_2$ instead of~\eqref{sup}, where $\hat{a}_{1,2}\equiv\hat{a}_{{1,2}}(x,t)$ are field operators which correspond to mean-field amplitudes $\Psi_{1,2}(x,t)$ defined in~\eqref{sup}. We assume that experimental conditions allow the formation of atomic bright solitons in each well of harmonic double-well potential. In particular, these conditions may be realized be means of manipulation with weak trapping potential $V(x)$. In an experiment this manipulation may be performed with the help of a dipole trap and laser field. 
 
Then, considering linear superposition state like~\eqref{ham} we can write the total Hamiltonian $\hat{H}$ for two BEC solitons in the form 
\begin{subequations}\label{ham}
\begin{equation}
\hat{H}=\hat{H}_1+\hat{H}_2+\hat{H}_{int},
\end{equation}
where $\hat{H}_{j}$ ($j=1,2$) is the Hamiltonian for condensate particles in $j$-th well; while $\hat{H}_{int}$ accounts the coupling between two wells due to the soliton overlapping.
In the second quantization form we explicitly have
\begin{eqnarray}
&&\hat{H}_{j} = \int_{-\infty}^\infty \hat{a}_j^\dag\left(-\frac{1}{2}\frac{\partial^2}{\partial x^2}\right)\hat{a}_{j}dx;\\
&&\hat{H}_{int} = -\frac{u}{2}\int_{-\infty}^\infty \Big(\hat{a}_1^\dag+\hat{a}_2^\dag\Big)^2\Big(\hat{a}_1+\hat{a}_2\Big)^2dx.
\end{eqnarray}
\end{subequations}
%Here, $N$ is an average total number of particles; $u>0$ is an absolute value of Kerr-like nonlinearity that appear due to atom-atom scattering in Born-approximamtion.
%In practice bright matter wave solitons are obtained for atomic condensates with the negative scattering length that corresponds to attractive particles~\cite{Strecker2002,Strecker2002a}, that is taken into account in~\eqref{ham}. 

The annihilation (creation) operators of bosonic fields denoted as $\hat{a}_{j}$ (${\hat{a}^\dag_{j}}$) with $j=1,2$ obey the commutation relations:
\begin{equation}
[\hat{a}_i(x), {\hat{a}^\dag_{j}(x')}] = \delta (x-x')\,\delta_{ij}; \quad i,j =1,2.
\label{commutator}
\end{equation}

In the Hartree approximation for a large particle number, $N>>1$, one can assume that the quantum $N$-particle two-soliton state is the product of $N$ two-soliton states and can be written as~\cite{Haus1989,Haus1989a, Alodjants1995}
%Physically, this assumption is valid for BECs taken in equilibrium at zero temperature. 
%Thus, the collective ground state for the whole system 
\begin{equation}\label{GS}
\left|\Psi_N\right\rangle = \frac{1}{\sqrt{N!}}\left[\int_{-\infty}^\infty\left(\Psi_1(x,t)\hat{a}_1^\dag e^{-i\beta_1t} + \Psi_2(x,t)\hat{a}_2^\dag e^{-i\beta_2t}\right)dx\right]^N\left|0\right\rangle,
\end{equation}
where $\Psi_j(x,t)$ are unknown wave functions, $|0\rangle\equiv|0\rangle_1 |0\rangle_2$ is a two-mode vacuum state. The state~\eqref{GS} is normalized as $\left\langle\Psi_N\big|\Psi_N\right\rangle = 1$, and
%Note that the state vector shown in Eq.~\eqref{GS} relates to the Hartree approach for bosonic systems~\cite{Haus1989,Haus1989a, Alodjants1995}, which is valid for a large number of particles $N$.
the bosonic field-operators $\hat{a}_{j}$ act on it as% the state~\eqref{GS} as
\begin{equation}
\hat{a}_j\left|\Psi_N\right\rangle = \sqrt{N}\Psi_j(x,t)e^{-i\beta_jt}\left|\Psi_{N-1}\right\rangle.
\end{equation}
Applying variational field theory approach based on the ansatz $\Psi_j(x,t)$ we obtain the Lagrangian density in the form:%~\cite{raghavan}:
\begin{equation}\label{lag_0}
L_0 = \frac{1}{2} \sum_{j=1}^2\left(i\left[\Psi_j^*\dot{\Psi}_j - \dot{\Psi}_j^*\Psi_j\right] - \left|\frac{\partial\Psi_j}{\partial x}\right|^2\right) + \frac{uN}{2}\left(\Psi_1^*e^{i\beta_1t}+\Psi_2^*e^{i\beta_2t}\right)^2\left(\Psi_1e^{-i\beta_1t}+\Psi_2e^{-2i\beta_2t}\right)^2,
\end{equation}
where we suppose $N-1\approx N$ and omit the common term $N$.

Noteworthy, from~\eqref{lag_0} one can obtain the coupled GPEs for $\Psi_j$-functions as 

\begin{subequations}\label{GP0}
\begin{eqnarray}
%i\frac{\partial}{\partial t}\Psi_j&=&-\frac{1}{2}\frac{\partial^2}{\partial x^2}\Psi_j + u(N-1)\left(\Psi_1^*e^{i\beta_1t}+\Psi_2^*e^{i\beta_2t}\right)\left(\Psi_1e^{-i\beta_1t}+\Psi_2e^{-i\beta_2t}\right)^2e^{i\beta_jt}.\\
i\frac{\partial}{\partial t}\Psi_1&=&-\frac{1}{2}\frac{\partial^2}{\partial x^2}\Psi_1 - uN\left(\left|\Psi_1\right|^2 + 2\left|\Psi_2\right|^2\right)\Psi_1\\ 
&-&uN\left(\left|\Psi_2\right|^2 + 2\left|\Psi_1\right|^2\right)\Psi_2e^{-i\beta t} - uN\Psi_1^*\Psi_2^2e^{-2i\beta t} - uN\Psi_2^*\Psi_1^2e^{i\beta t};\nonumber\\
i\frac{\partial}{\partial t}\Psi_2&=&-\frac{1}{2}\frac{\partial^2}{\partial x^2}\Psi_2 - uN\left(\left|\Psi_2\right|^2 + 2\left|\Psi_1\right|^2\right)\Psi_2\\ 
&-&uN\left(\left|\Psi_1\right|^2 + 2\left|\Psi_2\right|^2\right)\Psi_1e^{i\beta t} - uN\Psi_2^*\Psi_1^2e^{2i\beta t} - uN\Psi_1^*\Psi_2^2e^{-2i\beta t}\nonumber,
\end{eqnarray}
\end{subequations}
where $\beta=\beta_2-\beta_1$ is the energy (frequency) spacing. 

Set of Eqs.~\eqref{GP0} lead to the known problem for transitions between two lowest self-trapped states of condensates in the nonlinear coupled mode approach if we account~\eqref{wf1} for $\Psi_j(x,t)$ condensate wave functions representation~\cite{Ostrovskaya2000,Elyutin2001}. 

On the other hand, Eqs.~\eqref{GP0} can be recognized in the framework of soliton interaction problem that may be solved by means of perturbation theory for solitons~\cite{Karpman1981}. In particular, in accordance with Karpman's approach we can recognize in~\eqref{GP0} terms proportional to $\epsilon_{jk}=\Psi_j^*\Psi_k^2 + 2|\Psi_j|^2\Psi_k$, $j,k=1,2$, $j\neq k$ as perturbations for two fundamental bright soliton solutions. Physically, $\epsilon_{jk}$ implies the nonlinear Josephson coupling between the solitons. 

In this work we establish a variational approach for solution of Eqs.~\eqref{GP0}, cf.~\cite{Tsarev2019}. For the weakly coupled condensate states, i.e. for $\epsilon_{jk}\simeq0$, set of Eqs.~\eqref{GP0} reduces to two independent GPEs:
\begin{equation}\label{GP1}
i\frac{\partial}{\partial t}\Psi_j = -\frac{1}{2}\frac{\partial^2}{\partial x^2}\Psi_j - uN\left|\Psi_j\right|^2\Psi_j,
\end{equation}
which possess bright (non-moving) soliton solutions 
%which of Eqs.~\eqref{GP1} which looks like
\begin{equation}\label{anz_00}
\Psi_j(x,t)=\frac{\sqrt{uN_j}}{2}\sech\left[\frac{uN_j}{2}x\right]e^{i\frac{u^2N_j^2}{8}t}.
\end{equation}
%where $\theta_j = \frac{u^2N_j^2}{8}t$ is $j$-th soliton phase. 

In the case of $\epsilon_{jk}\neq0$ and for non-zero inter-soliton distance $\delta$, we examine ansatzes for $\Psi_j(x,t)$ in the form
\begin{subequations}\label{anz_0}
\begin{eqnarray}
 \Psi_1(x,t)&=&\frac{\sqrt{uN_1}}{2}\sech\left[\frac{uN_1}{2}x - \delta\right]e^{i\theta_1};\\
 \Psi_2(x,t)&=&\frac{\sqrt{uN_2}}{2}\sech\left[\frac{uN_2}{2}x + \delta\right]e^{i\theta_2}.
\end{eqnarray}
\end{subequations}
In particular, our approach presumes the existence of two well distinguished solitons (separated by the small distance $\delta$, with the shape preserved) interacting through dynamical variation of the particle numbers ($N_j\equiv N_j(t)$) and phases ($\theta_j\equiv\theta_j(t)$), which occurs in the presence of weak coupling between the solitons. In other words, $N_j$ and $\theta_j$ should be considered as time-dependent (variational) parameters. 
Setting~\eqref{anz_0} in~\eqref{lag_0} we obtain
\begin{equation}\label{lag}
L = \int_{-\infty}^\infty L_0 dx\simeq - z\dot{\theta} + \Lambda z^2 + \frac{\Lambda}{2}\left(1 - z^2\right)^2I(z,\Delta)\left(\cos[2\Theta]+2\right) + \Lambda\left(1 - z^2\right)J(z,\Delta)\cos[\Theta],
\end{equation}
where $z=(N_2-N_1)/N$ ($N_{1,2}=\frac{N}{2}(1\mp z)$) is the particle number population imbalance; $\Theta=\theta_2-\theta_1-(\beta_2-\beta_1)t \equiv \theta - \Omega t$ is an effective time-dependent phase-shift between the solitons. 

Physically, $\Omega$ is an angular frequency spacing between the ground and first excited macroscopic states of the condensate; it represents a vital (measured) parameter for metrological purposes in this work. 
In~\eqref{lag} we introduce the notation $\Lambda=N^2u^2/16$ and define the functionals 
\begin{subequations}\label{funcs}
\begin{eqnarray}
I\equiv I(z,\Delta)&=&\int_{-\infty}^\infty\sech^2\left[\left(1-z\right)\left(x-\Delta\right)\right]\sech^2\left[\left(1+z\right)\left(x+\Delta\right)\right]dx;\\
J\equiv J(z,\Delta)&=&\sum_{s=\pm1}\left(\int_{-\infty}^\infty\left(1+sz\right)^2\sech^3\left[\left(1+sz\right)\left(x+s\Delta\right)\right]\sech\left[\left(1-sz\right)\left(x-s\Delta\right)\right]\right),\nonumber\\
\end{eqnarray}
\end{subequations}
where $\Delta\equiv~\frac{Nu}{4}\delta$ is a normalized distance between solitons. 

Finally, by using Eq.~\eqref{lag} for the population imbalance and phase-shift difference, $z$ and $\Theta$, we obtain the set of equations
\begin{subequations}\label{eqs}
\begin{eqnarray}
\dot{z}&=&\left(1-z^2\right)\left\{\left(1-z^2\right)I\sin[2\Theta] + J\sin[\Theta]\right\};\label{eqs_a}\\
\dot{\Theta}&=& - \frac{\Omega}{\Lambda} + 2z + \frac{\rm{d}}{\rm{d}z}\left\{\frac{1}{2}\left(1-z^2\right)^2I\left(\cos[2\Theta]+2\right) + \left(1-z^2\right)J\cos[\Theta]\right\},\label{eqs_b}
\end{eqnarray}
\end{subequations}
where dots denote derivatives with respect to renormalized time $\tau=\Lambda t$. 

In contrast to the problem with coupled Gaussian-shape condensates (cf.~\cite{Ostrovskaya2000,Elyutin2001}), the solutions of Eqs.~\eqref{eqs} crucially depend on features of governing functionals $I(z,\Delta)$ and $J(z,\Delta)$. In Appendix we represent some analytical approximations for $I(z,\Delta)$ and $J(z,\Delta)$, which we exploit further.

\section{Steady-state (SS) solutions}

%The set of equations~\eqref{eqs} cannot be integrated analytically, due to~\eqref{funcs}, whose functional dependence on $z$ and $\Delta$ is through an integral.

\subsection{SS solution for $z^2=1$}

The SS solutions of~\eqref{eqs} play a crucial role for metrological purposes with coupled solitons cf.~\cite{Tsarev2018}. 
%, $\Theta=0$ or $\Theta=\pi$. Further analysis of~\eqref{eqs_b} will give the expression of the missing parameter, $\Theta$ or $z$, respectively.
We start from the SS solution $z^2=1$ of Eq.~\eqref{eqs_a} by setting the time-derivatives to zero. As seen from~\eqref{funcs}, in the limit of maximal population imbalance, $z^2=1$, $I$ and $J$ do not depend on $\Delta$ and approach
\begin{subequations}\label{func_approx_N00N}
\begin{eqnarray}
I(z,\Delta)&=& 1;\\
J(z,\Delta)&=&\pi.
\end{eqnarray}
\end{subequations}
Substituting $z^2=1$ and~\eqref{func_approx_N00N} into~\eqref{eqs_b} we obtain
%The Eqs.~\eqref{eqs} simplify significantly in this limit:
%\begin{subequations}
%\begin{eqnarray}
%\dot{z}&=&0;\\
%\dot{\Theta}&=& - \frac{\Omega}{\Lambda} + 2z - 2zJ\cos[\Theta] = - \frac{\Omega}{\Lambda} + 2z\left(1 - \pi\cos[\Theta]\right) = 0,
%\dot{\Theta}&=&- \frac{\Omega}{\Lambda} + 2z\left(1 - \pi\cos[\Theta]\right) = 0,
%\end{eqnarray}
%\end{subequations}
%i.e.
\begin{subequations}\label{n00n_sol}
\begin{eqnarray}
 z^2&=&1;\\
 \Theta&=&\arccos\left[\frac{2\Lambda - \sign[z]\Omega}{2\pi\Lambda}\right].\label{N00N_b}
\end{eqnarray}
\end{subequations}
Notably, in quantum domain the SS solutions~\eqref{n00n_sol} admit the existence of quantum states with maximal population imbalance $z= \pm1$ and phase difference. The latter depends on the frequency spacing $\Omega$, which is the subject of precise measurement with maximally path-entangled $N00N$-states in this paper.

Below we perform the analysis of the SS solutions of Eqs.~\eqref{eqs} in two limiting cases $\Omega\neq0$, $\Delta\simeq0$ and $\Omega\simeq0$, $\Delta\neq0$.

\subsection{SS solutions for $\Theta=0,\pi$ and $\Delta\simeq0$}

To find the SS solutions we rewrite~\eqref{eqs_b} as
\begin{equation}\label{z_Delta_0}
 \frac{\Omega}{\Lambda} = 2z - 6z\left(1-z^2\right)I + \frac{3}{2}\left(1-z^2\right)^2\frac{\partial I}{\partial z} - 2zJ + \left(1-z^2\right)\frac{\partial J}{\partial z}
\end{equation}
for $\Theta=0$ and
\begin{equation}\label{z_Delta_pi}
 \frac{\Omega}{\Lambda} = 2z - 6z\left(1-z^2\right)I + \frac{3}{2}\left(1-z^2\right)^2\frac{\partial I}{\partial z} + 2zJ - \left(1-z^2\right)\frac{\partial J}{\partial z}
\end{equation}
for $\Theta=\pi$, respectively.

In Appendix we represent a polynomial approximation for $I,J$ functionals~\eqref{funcs}. Since the equations obtained from~\eqref{z_Delta_0} and~\eqref{z_Delta_pi} are quite cumbersome, here we just briefly analyze the results. 

In the limit of closely spaced solitons and $\Theta=0$, the population imbalance $z$ at equilibrium depends only on $\Omega$ and obeys
\begin{equation}
 %\Omega = 1.08z^7- 7.8z^5 + 14.88z^3 - 12.44z.\label
 \frac{\Omega}{\Lambda} = 1.2 z^7 - 8 z^5 + 15 z^3 - 12.5 z.\label{Omega_0}
\end{equation}
Similarly, for fixed soliton phase difference $\Theta=\pi$ we have 
\begin{equation}
 %\Omega = 1.08z^7 - 3z^5 + 12.24z^3 - 2.04z\label{Omega_p}.
 \frac{\Omega}{\Lambda} = 1.2 z^7 - 3.2 z^5 + 12.3 z^3 - 2 z.\label{Omega_p}
\end{equation}
%The numerical simulation reviled that no more stationary states exist for the current model, see Sec. IV.

\begin{figure}[h!]
\begin{center}
\includegraphics[width=0.75\linewidth]{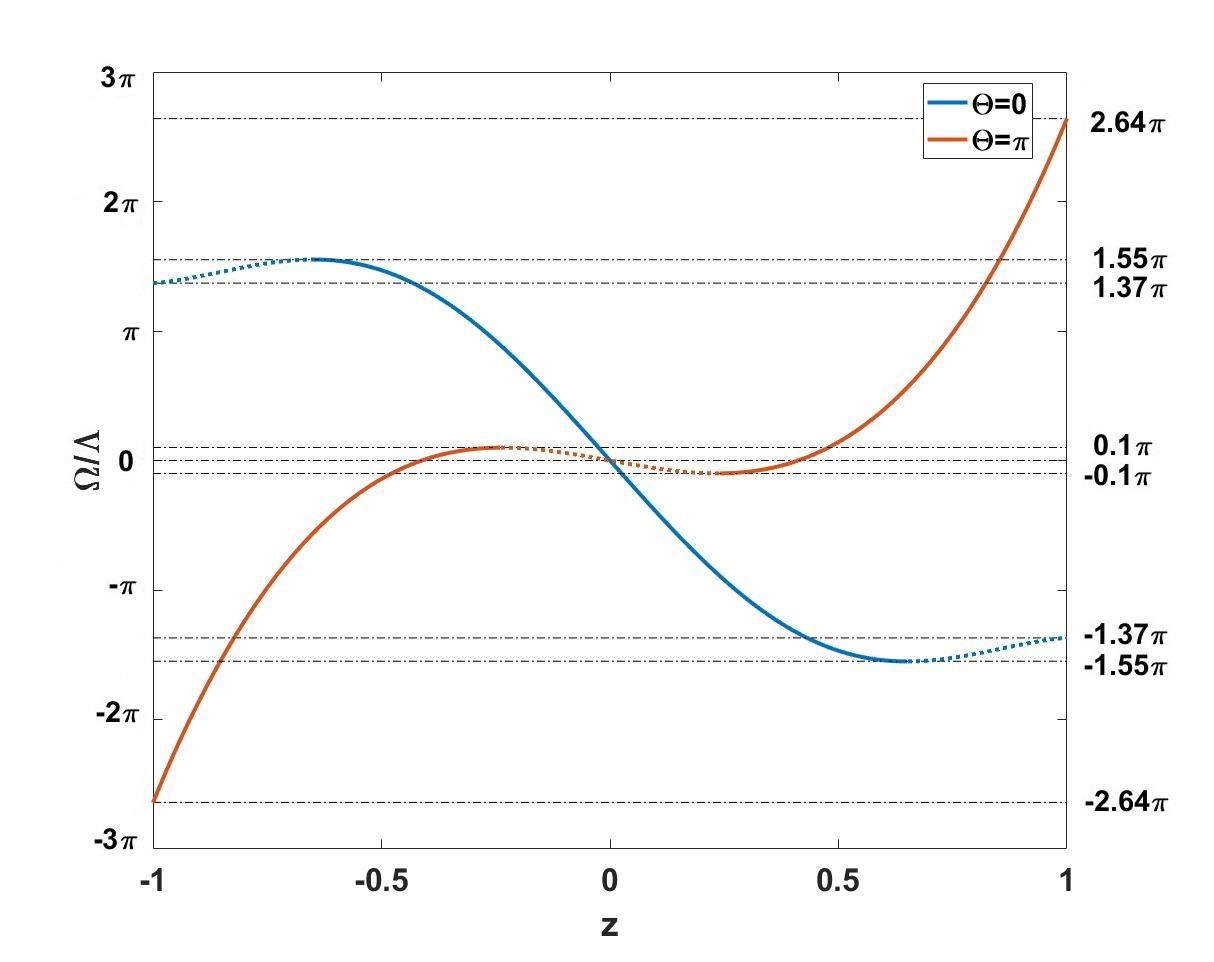}
\end{center}
\caption{Normalized frequency spacing $\Omega/\Lambda$ (dashed lines) versus reduced population imbalance $z$ for Eq.~\eqref{Omega_0} (blue line) and Eq.~\eqref{Omega_p} (red line), respectively. Dashed regions correspond to unstable solutions. }
\label{FIG:Omega}
\end{figure}

We plot the graphical solutions of Eqs.~\eqref{Omega_0},~\eqref{Omega_p} in Fig.~\ref{FIG:Omega}; blue and red curves characterize the right parts of Eqs.~\eqref{Omega_0},~\eqref{Omega_p}, respectively. Straight lines in Fig.~\ref{FIG:Omega} correspond to different values of $\Omega/\Lambda$ ratio. These lines cross the curves in the points which indicate the solutions of Eqs.~\eqref{Omega_0},~\eqref{Omega_p}. Notice, that the solid blue and red curves denote the values of $\Omega/\Lambda$ and $z$ corresponding to the stable SS solutions, while the dotted ones describe parametric unstable solutions. As seen from Fig.~\ref{FIG:Omega} at phase difference $\Theta=0$ there exists one stable SS solution for any $z\in[-0.7;0.7]$ and only unstable solutions for $|z|>0.7$. At $|\Omega/\Lambda|>1.55\pi$ no SS solutions exist.

On the other hand, at $\Theta=\pi$ there exists a tiny region $-0.1\pi\leq\Omega/\Lambda\leq0.1\pi$ possessing two SS solutions simultaneously. One stable SS solution exists within the domain $0.1\pi<|\Omega/\Lambda|\leq2.64\pi$.

\subsection{SS solutions for $\Theta=0,\pi$ and $\Omega\simeq0$}

At $\Omega=0$ Eqs.~\eqref{Omega_0},~\eqref{Omega_p} admit the SS solutions, which look like:
\begin{subequations}\label{STst}
\begin{eqnarray}
z=0,&\Theta=0;\label{STst_a}\\
z=0,&\Theta=\pi;\label{STst_b}\\
z^2\approx0.17&\Theta=\pi.\label{STst_c}
\end{eqnarray}
\end{subequations}
%The solutions~\eqref{STst_a} and~\eqref{STst_c} are stationary, while~\eqref{STst_b} is parametric unstable.
%Noteworthy, the SS states like in~\eqref{STst_c} are degenerate, $z =\pm z_0\approx\pm\ 0.41$ and determine some finite fraction of population in the solitons. This circumstance allows us to form another type of entangled superposition state that is so-called Schrodinger Cat state (SC-state). 
%Let us now investigate the role of $\Delta$ in solitons dynamics. 
As seen from 
%As mentioned earlier at $\Delta\approx0$, $\Omega\simeq0$, and $\Theta=\pi$ 
Eq.~\eqref{STst_b}, at relative phase $\Theta=\pi$~\eqref{z_Delta_pi} possesses three solutions: a parametrically unstable solution occurs at $z=0$ and two degenerate SS solutions appear for $z=\pm z_0$. $z_0$ varies from $0.41$ at $\Delta\approx0$ to $0.64$ at $\Delta\approx2.8$ for non-zero soliton inter-distance, respectively. For $\Delta > 2.8$ these SS solutions do not exist.

\begin{figure*}[h!]
\center{\includegraphics[width=0.75\linewidth]{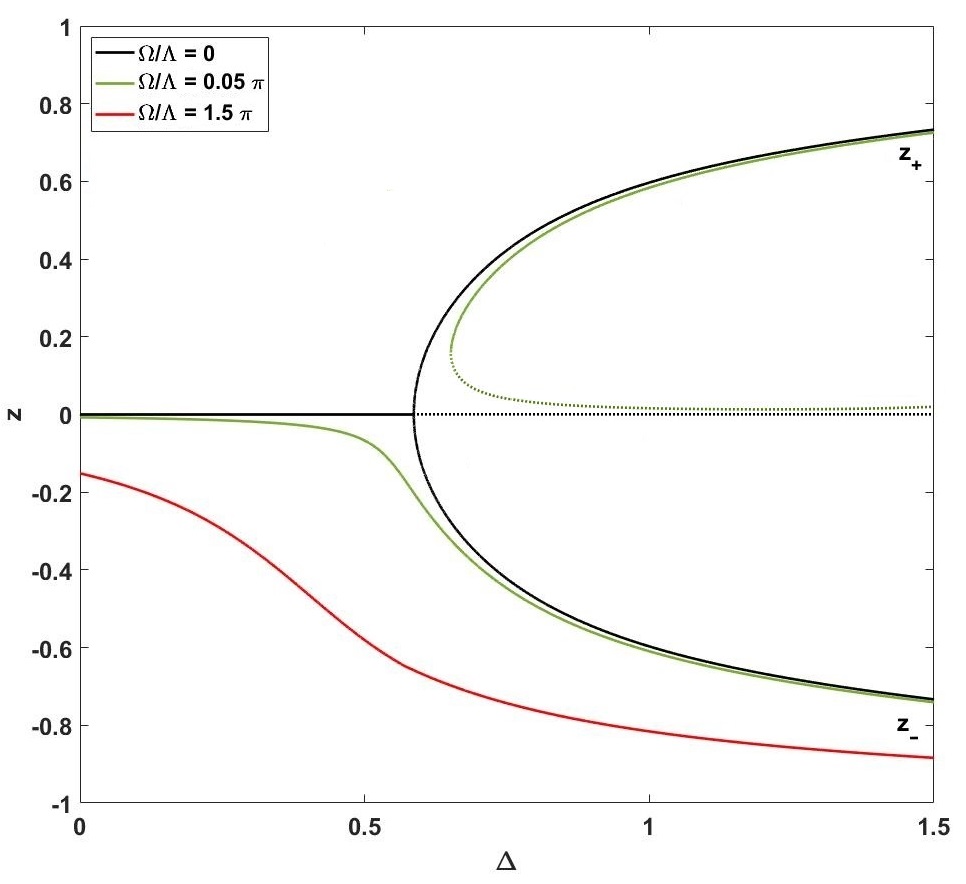}}
\caption{Population imbalance $z$ versus $\Delta$ for $\Theta=0$ and different $\Omega$. Solid curves denote SS solutions of~\eqref{z_Delta_0} and the dotted line represents (parametrically) unstable solutions.}
\label{FIG:D(z)}
\end{figure*}

In Fig.~\ref{FIG:D(z)} we represent the more general analysis of SS solutions for $\Theta=0$ as functions of inter-soliton distance $\Delta$ for different $\Omega$. For that we exploit the sixth-order polynomial approximation, see Appendix. In particular, at $\Omega\simeq0$ there exists one solution at $z=0$, stable at $\Delta\leq\Delta_c\approx0.5867$. For $\Delta>\Delta_c$ this solution becomes parametrically unstable. 

On the other hand, for $\Delta>\Delta_c$ Eqs.~\eqref{z_Delta_0} possesses the degenerate SS solutions similar to the ones at $\Theta=\pi$. The bifurcation for population imbalance $z$ occurs at $\Delta=\Delta_c$; in Fig.~\ref{FIG:D(z)} the $z_+$ (upper,positive) and $z_-$ (lower, negative) branches characterize this bifurcation. 
%Fig.~\ref{FIG:D(z)} demonstrates SS solution for $z$ as function of $\Delta$ at different $\Omega/\Lambda>0$. Contrary, for $\Omega/\Lambda<0$ the curves in Fig.~\ref{FIG:D(z)} are flipped vertically. 
%From Fig.~\ref{FIG:D(z)} clearly seen bifurcation behavior for population imbalance $z$ that occurs in the point $\Delta=\Delta_c$ at $\Omega=0$. For $\Delta>\Delta_c\approx0.5867$ the stationary solution degenerates, $z=\pm z_0$, and \textbf{ solutions of~\eqref{z_Delta_0} become parametrically unstable.} 
In the vicinity of $\Delta_c$ we can consider $z_\pm=\pm z_0$, where%dependence on $\Delta$ is 
\begin{equation}\label{z0_Tailor}
 %z_0=1.2\sqrt{\Delta - 0.5867}.
 %z_0=10.6(\Delta - 0.5867).
 z_0=1.2\sqrt{\Delta - \Delta_c}.
\end{equation}
%\textbf{As shown further, the oscillations occur in the vicinity of these stationary points and $z=0$, plotted in Fig.~\ref{FIG:D(z)} by the dotted line, plays the role of separatrix.} 

At $\Omega\neq0$ the behavior of SS solutions depending on distance $\Delta$ complicates - see green curves in Fig.~\ref{FIG:D(z)}. The solid curves correspond to SS solutions for different $\Delta$, while the dotted ones describe the unstable solutions. From Fig.~\ref{FIG:D(z)} it is clearly seen that for $\left|\Omega\right|>0$ there is no bifurcation for population imbalance $z$ and two stationary solution branches $z_{\pm}$ occur with $|z_-|>|z_+|$. 

At relatively large values of parameter $\Omega/\Lambda$ only one SS solution exists - see red curve in Fig.~\ref{FIG:D(z)}. 

%\textbf{As before, there is the third, unstable solution being the separatrix between two oscillation areas, see green curves in Fig.~\ref{FIG:D(z)}. For $\Omega/\Lambda>0.5\pi$ the upper branch disappear and the only stationary solution is at $z<0$, see the red curve in Fig.~\ref{FIG:D(z)}.}

%\textbf{The function $z(\Delta)$ at $\Theta=\pi$ is much simpler. At $\left|\Omega/\Lambda\right|<0.1\pi$ the stationary solution is degenerate, $z=\pm z_0$, with the separatrix $z\approx0$. For $\Omega/\Lambda>0.1\pi$ the solution $z<0$ disappears and only $z>0$ solution remains (for negative $\Omega$ $z<0$ solution disappears). We see no need to give any figure here, the reader can find these solutions on the phase diagrams $z-\Theta$ in the next Section}.

\section{Mean-field dynamics}

\subsection{Small amplitude oscillations}

We start our analysis here from small amplitude oscillations close to the SS solutions~\eqref{STst}. 
% Continuing to study the dynamics of solitons in the mean-field approximation, it is necessary to investigate the stability of the stationary solutions obtained. 
For that we linearize Eqs.~\eqref{eqs} in the vicinity of~\eqref{STst}, assuming $0\leq\Delta<0.6$ and $\Omega<<1$. The first assumption allows to use the approximation of $I,J$-functionals by the fourth-degree polynomial, see Appendix. 
%The second one is necessary, since~\eqref{STst} were obtained for $\Omega=0$, so they are valid only in the vicinity of this value. 

For zero-phase oscillations, i.e. for $\Theta\approx0$ ($\cos\left[\Theta\right]\approx1$, $\sin\left[\Theta\right]\approx\Theta$), from~\eqref{eqs} we obtain 
\begin{equation}\label{2ODE}
%\ddot{z}\approx- z(-179\Delta^2 - 44\Delta + 67).
\ddot{z} + \omega_0^2(\Delta)z = f_0(\Delta)\Omega
\end{equation}
with the solution
\begin{equation}\label{lin_osc}
 %z(\tau) = A\cos\left[\omega\tau\right] + B(\Delta)\Omega,
 %z(\tau) = \left(z(0)-\frac{f}{\omega^2}\Omega\right)\cos\left[\omega\tau\right] + \frac{f}{\omega^2}\Omega,
 z(\tau) = A\cos\left[\omega_0\tau\right] - \Omega\frac{f_0}{\omega_{0}^2},
\end{equation}
where $A$ and $\omega_0(\Delta)=13.4\sqrt{0.37-\Delta^2 - 0.25\Delta}$ are the amplitude and angular frequency of oscillations, respectively. The last term in~\eqref{lin_osc} with $f_0(\Delta)=5.36-0.8\Delta-4.22\Delta^2$ plays the role of constant "external downward displacement force" that vanishes at $\Omega\simeq0$. Notably, at $\Delta>0.5$ oscillations become anharmonic and $z(\tau)$ diverges at $\Delta>0.5867$. For $\Delta=0$ the frequency of oscillations approaches $\omega \approx 8.15$, that agrees with numerical solution of~\eqref{eqs}.

At $\Delta=\Delta_c\simeq0.5867$ the SS solution~\eqref{STst} splits into two degenerate solutions with $z=\pm z_0$ and $z_0$ determined by~\eqref{z0_Tailor}, see Fig.~\ref{FIG:D(z)}. Near these points the equation, similar to~\eqref{2ODE}, has a form
\begin{equation}
 \ddot{z} + \omega^2z = - 18\Delta_-\sqrt{\Delta_-} - f(\Delta_-)\Omega
\end{equation}
that implies a solution
\begin{equation}\label{2ODE_2}
 z(\tau) = \pm\left(1.2 - \frac{18\Delta_-}{\omega^2}\right)\sqrt{\Delta_-} + A\cos[\omega\tau] - \Omega\frac{f(\Delta_-)}{\omega^2},
\end{equation}
where $\Delta_-\equiv\Delta-\Delta_c$, 
% is detuning deviation at which SS solution occurs, 
$\omega = 14.53\sqrt{\Delta_{-} - 4.48\Delta_-^2 + 17.8\Delta_-^3 - 53.5\Delta_-^4}$ is the angular frequency of oscillations, $f = 3.4 - 7.26\Delta_- + 11\Delta_-^2$ is the "external" force. 
%To be more specific we give here some numbers. For example, at $\Delta=0.6$, which gives $\Delta'=0.0133$, $f \approx 3.3$ and $\omega_{0.6}\approx1.63$. Thus 
%\begin{equation}\label{2ODE_3}
 % z = \pm0.138 - 1.24\Omega + A\cos[\omega_{0.6}\tau]
%\end{equation}
%Another example: at $\Delta=0.75$, which gives $\Delta'=0.1633$, $f \approx 2.5$ and $\omega_{0.75}=4.19$, and so 
%\begin{equation}
 % z = \pm0.417 - 0.142\Omega + A\cos[\omega_{0.75}\tau]
%\end{equation}
The relative error for Eq.~\eqref{2ODE_2} is less than $5\%$. 
%As before, the term proportional to $f(\Delta')\Omega/\omega^2$, displaces the mean $z(\tau)$ to lower valuse causing the symmetry breaking, c.f. black and green curves in Fig.~\ref{FIG:D(z)}. The correction proportional to $18\Delta'/\omega^2$ increases the precision of~\eqref{z0_Tailor} for $\Delta>\Delta_c$.

%\textcolor{blue}{It is clearly seen from~\eqref{2ODE_2} and especially from~\eqref{2ODE_3}, that the SS solution with $z(\tau)>0$ do not exist for some $\Omega$. We have revealed the ratio of $\Delta'$ and $\Omega$ for which booth SS solutions exist for $\Theta=0$ and $\Delta>\Delta_c$. This ratio is
%\begin{equation}\label{ratio_Om_D}
 % \Omega<\frac{1230\Delta'^{7/2}\left(0.33 - \Delta'\right)}{\left(0.33 - \Delta'\right)^2 + 0.2}.\textrm{\textbf{Not correct!}}
%\end{equation}
%For example, at $\Omega=0.05\pi$~\eqref{ratio_Om_D} is held for $\Delta>0.64$, c.f. the upper solid green curve in Fig.~\ref{FIG:D(z)}.
%}

In the vicinity of SS points determined by Eq.~\eqref{STst_c}, we obtain $\pi$-phase oscillations characterized by 
\begin{equation}\label{self-trap}
 z(\tau) = \pm z_0 + A\cos\left[\omega_{\pi}\tau\right] + \frac{f_{\pi}}{\omega_{\pi}^2}\Omega
\end{equation}
with $\omega_{\pi} = \sqrt{2-0.9\Delta^2 - 0.3\Delta }$, $f_{\pi} = 0.1\left(\Delta^2 + 0.38\Delta + 5.5\right)$, and $z_0$ determined in~\eqref{STst_c}. For $\Omega\simeq0$ and $\Delta=0$ angular frequency is $\omega_{\pi} \approx 1.42$ that is much smaller than in zero-phase regime. 

%\begin{equation}
 %z(\tau) = \pm0.417 + \left(z(0)-\frac{f}{\omega^2}\Omega\right)\cos\left[\omega\tau\right],
 %z(\tau) = \pm0.42 + A(\Delta,\Omega)\cos\left[\omega\tau\right] + B(\Delta)\Omega,
% z(\tau) = \pm0.41 A\cos\left[\omega\tau\right],
%\end{equation}
%where $\omega = \sqrt{-0.9\Delta^2 - 0.3\Delta + 2}$. Again, for $\Omega<<1$ we can omit $f\Omega/\omega^2$ and obtain the oscillations in the form $z(\tau) = \pm0.417 + z_0\cos\left[\omega\tau\right]$. These oscillations occur with $\Delta$ in the mentioned domain; when $\Delta=0$, $\omega \approx 1.42$. 

\begin{figure*}[h!]
\begin{minipage}[h!]{0.49\linewidth}
\center{\includegraphics[width=\linewidth]{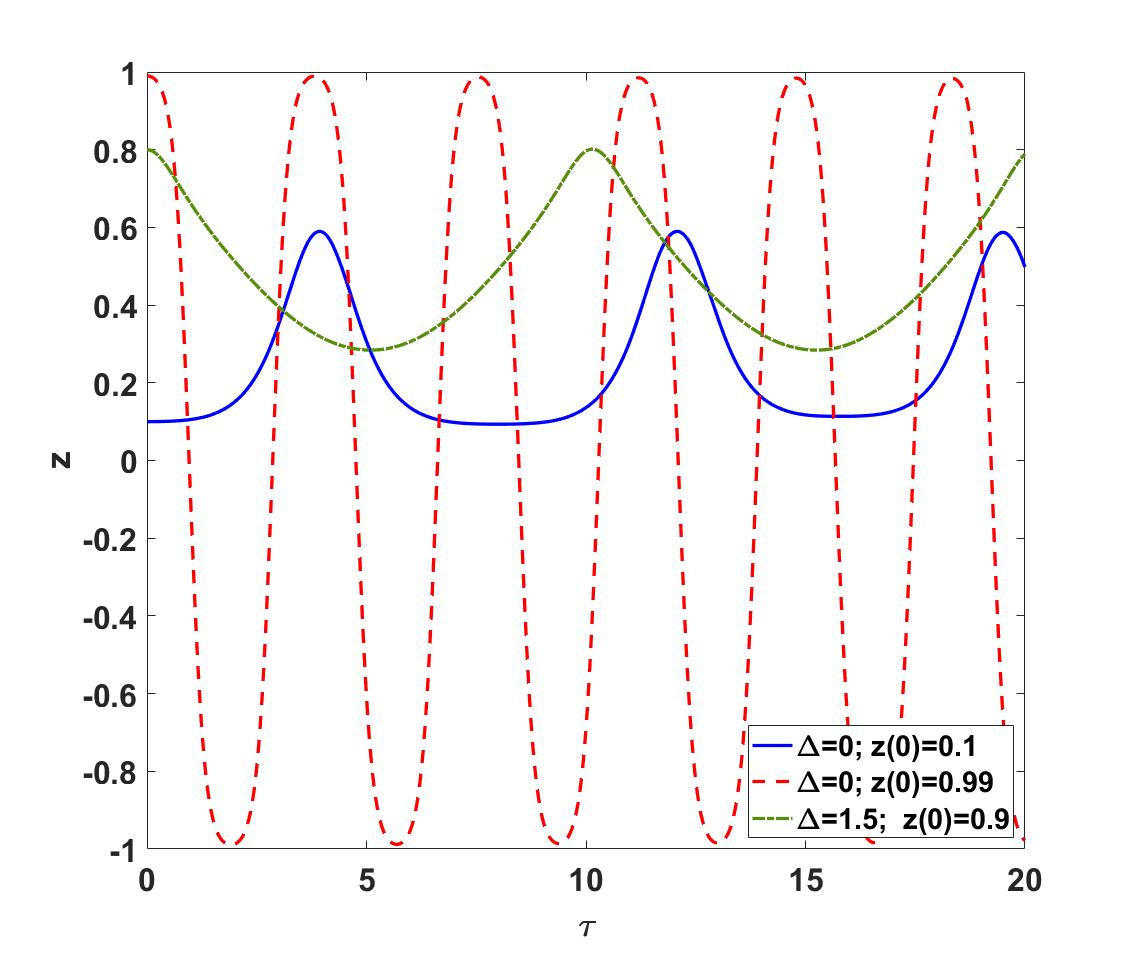} \\ a)}
\end{minipage}
\hfill
\begin{minipage}[h!]{0.49\linewidth}
\center{\includegraphics[width=\linewidth]{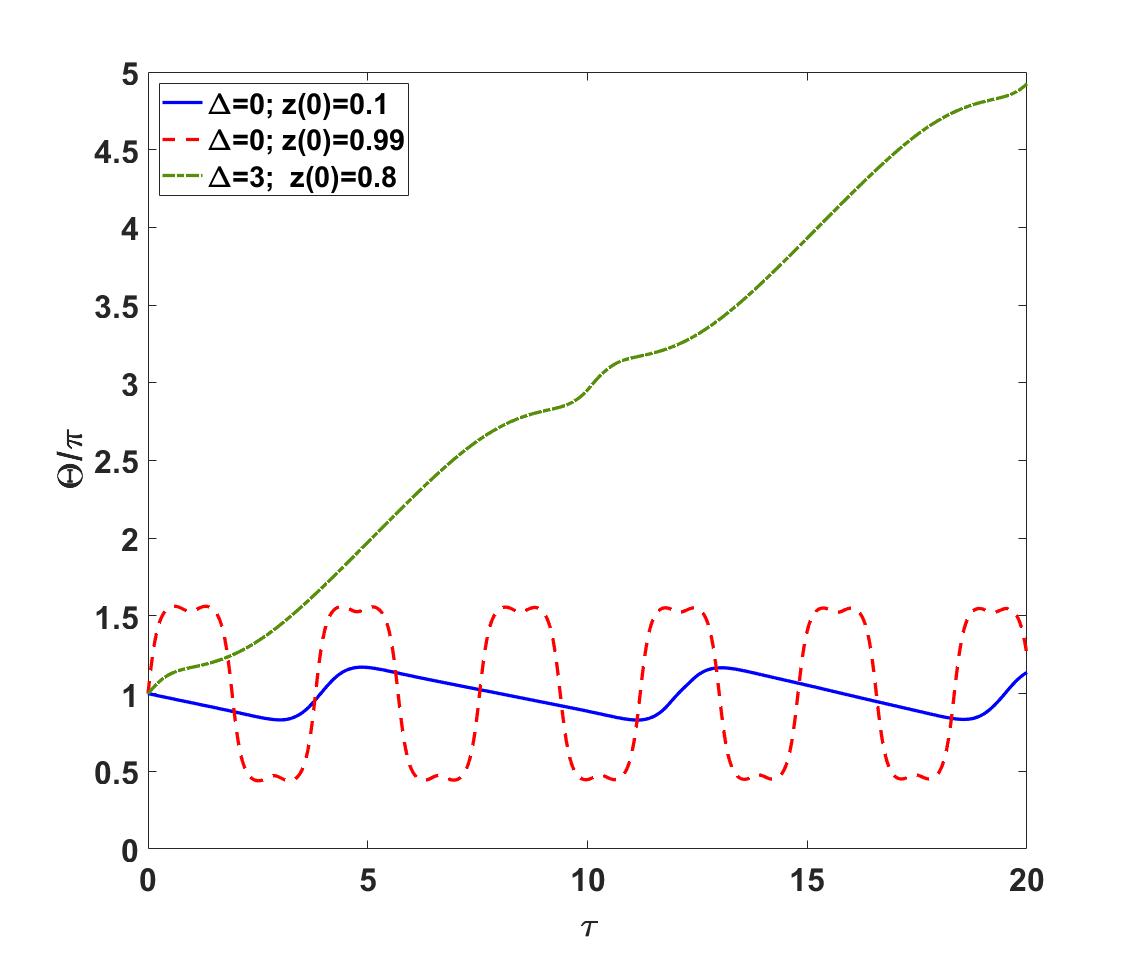} \\ b)}
\end{minipage}
%\center{\includegraphics[width=0.5\linewidth]{4_CR_NO.jpg}}
\caption{(a) the population imbalance $z(\tau)$ and (b) effective phase difference $\Theta(\tau)$ versus reduced time $\tau$ for $\Theta(0)=\pi$.}
\label{FIG:CR}
\end{figure*}

The analysis of~\eqref{eqs} in the vicinity of~\eqref{STst_b} reveals that this solution is parametric unstable, and highly nonlinear behavior is expected. Indeed, direct numerical simulation demonstrates anharmonic dynamics plotted in Fig.~\ref{FIG:CR}. For $0<|z|<0.5$ the nonlinear regime of self-trapping is observed, which turns into nonlinear oscillations at $|z|>0.5$. 

The analysis of SS solution~\eqref{n00n_sol} reveals the strong sensitivity to $z$-perturbation, when the condition $z^2=1$ is violated, the high-amplitude nonlinear oscillations occur. 
%\textcolor{red}{\hbox{Pls plot relevant curves in Fig.3}}. 
On the other hand, the solution~\eqref{n00n_sol} is robust to phase perturbations, which is an important property for metrology.

\subsection{Large separation limit, $\Delta>>1$}

For very large distance $\Delta$ between the solitons, $\Delta>>1$, the atom tunneling between them vanishes and the solitons become independent. Strictly speaking, in the limit $\Delta\rightarrow\infty$ the functionals $I,J\rightarrow0$ and Eqs.~\eqref{eqs} look like 
\begin{subequations}\label{large_Delta}
\begin{eqnarray}
\dot{z}&=&0;\\
\dot{\Theta}&=& - \Omega + 2z,
\end{eqnarray}
\end{subequations}
i.e. the population imbalance is a constant in time and the running-phase regime establishes.

For large but finite $\Delta$, SS solution $z=\pm z_0$ with $z_0\rightarrow1$ exists for the zero-phase regime, $\Theta=0$; for example, for $\Delta=10$ the SS population imbalance is $z_0\approx0.96$.

\subsection{Phase-space analysis}

\begin{figure*}[h!]
\begin{minipage}[h]{0.325\linewidth}
\center{\includegraphics[width=1\linewidth]{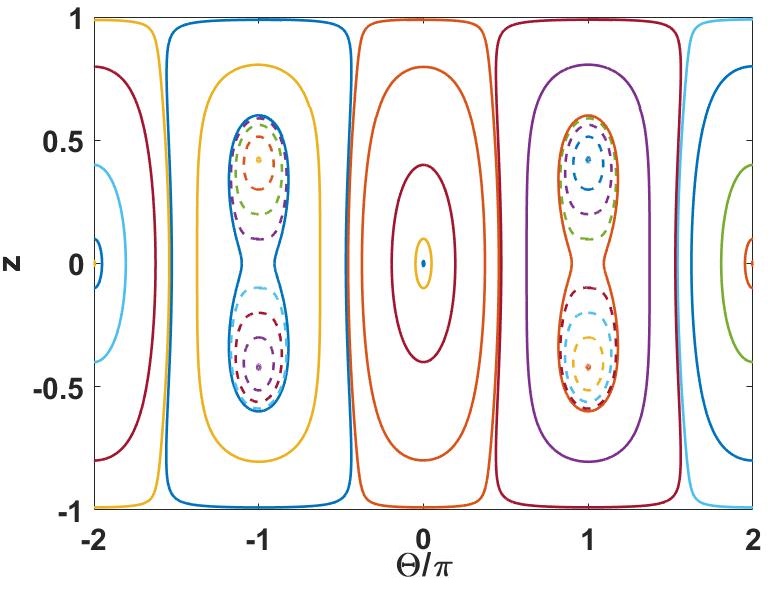} \\ a)}
\end{minipage}
\hfill
\begin{minipage}[h]{0.325\linewidth}
\center{\includegraphics[width=1\linewidth]{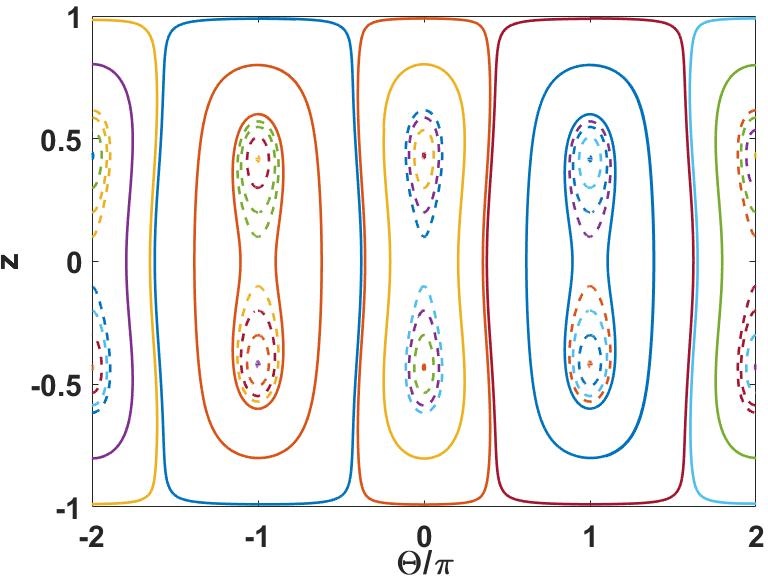} \\ b)}
\end{minipage}
\hfill
\begin{minipage}[h]{0.325\linewidth}
\center{\includegraphics[width=1\linewidth]{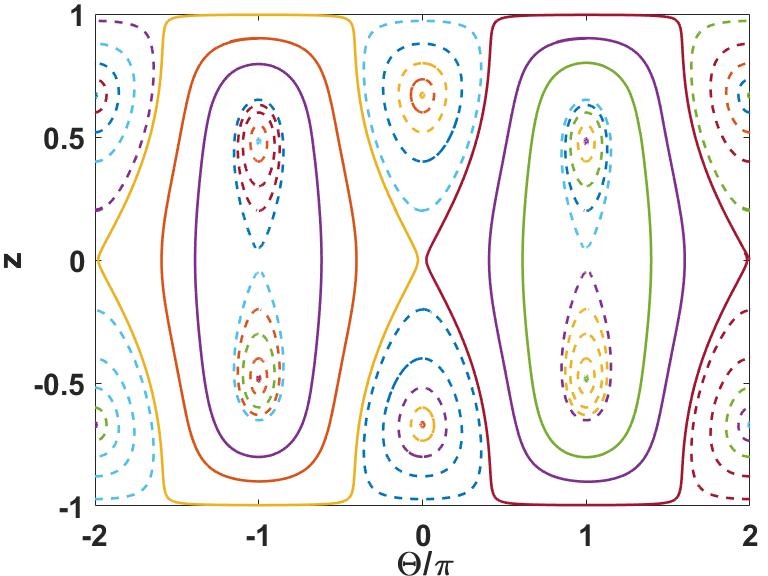} \\ c)}
\end{minipage}
\vfill
\begin{minipage}[h]{0.325\linewidth}
\center{\includegraphics[width=1\linewidth]{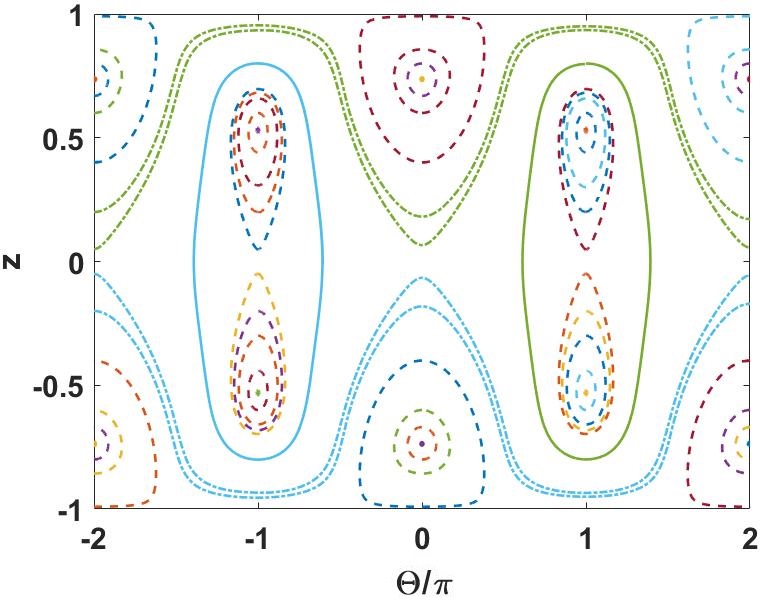} \\ d)}
\end{minipage}
\hfill
\begin{minipage}[h]{0.325\linewidth}
\center{\includegraphics[width=1\linewidth]{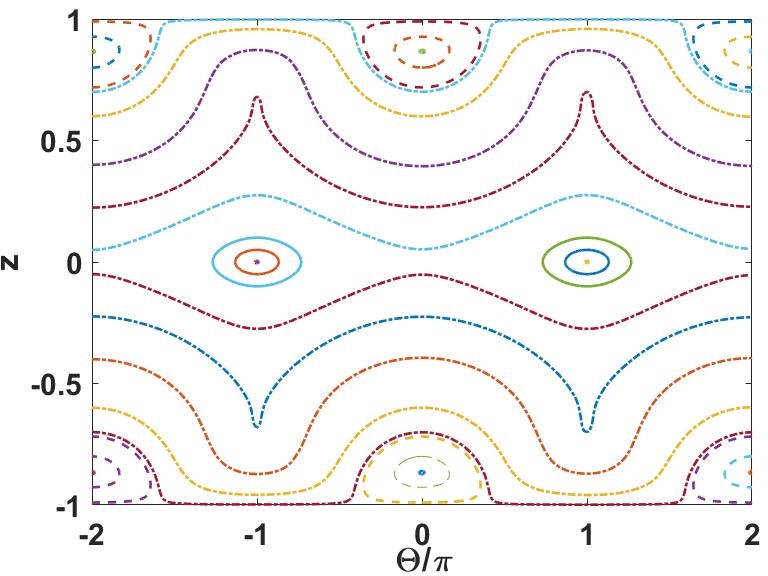} \\ e)}
\end{minipage}
\hfill
\begin{minipage}[h]{0.325\linewidth}
\center{\includegraphics[width=1\linewidth]{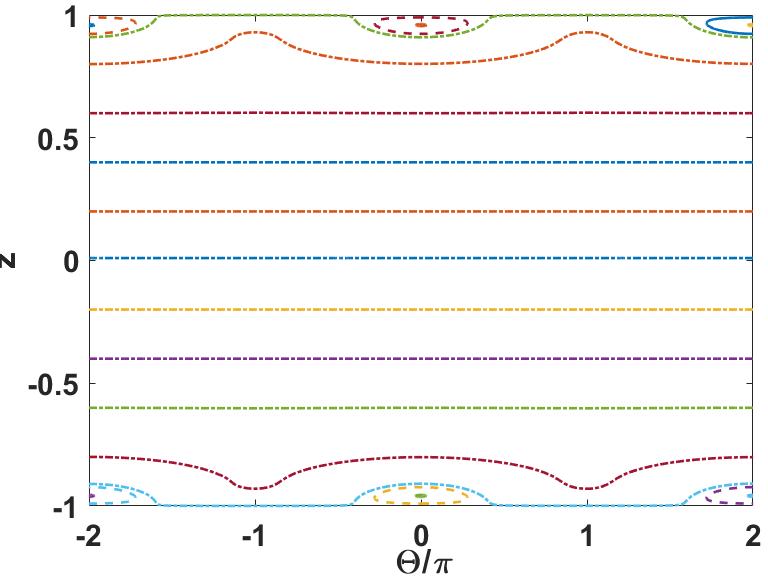} \\ f)}
\end{minipage}
\caption{Phase plane $z-\Theta$ at $\Omega=0$ and for (a) $\Delta = 0$; (b) $\Delta = 0.75$; (c) $\Delta = 1.2$; 
(d) $\Delta = 1.5$; (e) $\Delta = 3.0$; (f) $\Delta = 10$.}
\label{FIG:PD_Delta}
\end{figure*}

\begin{figure*}[h!]
\begin{minipage}[h]{0.325\linewidth}
\center{\includegraphics[width=1\linewidth]{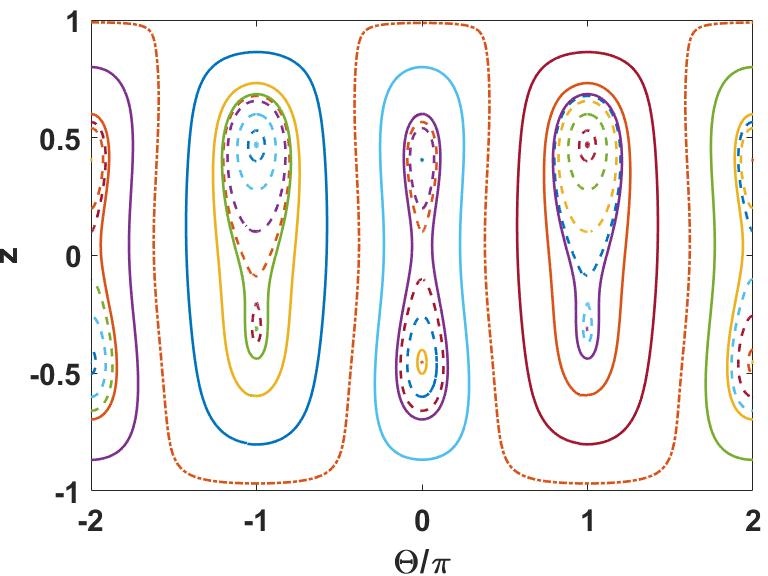} \\ a)}
\end{minipage}
\hfill
\begin{minipage}[h]{0.325\linewidth}
\center{\includegraphics[width=1\linewidth]{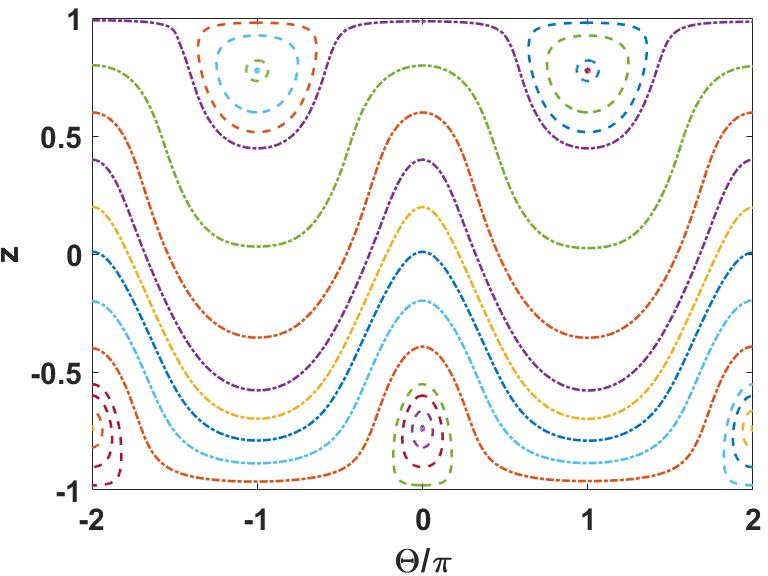} \\ b)}
\end{minipage}
\hfill
\begin{minipage}[h]{0.325\linewidth}
\center{\includegraphics[width=1\linewidth]{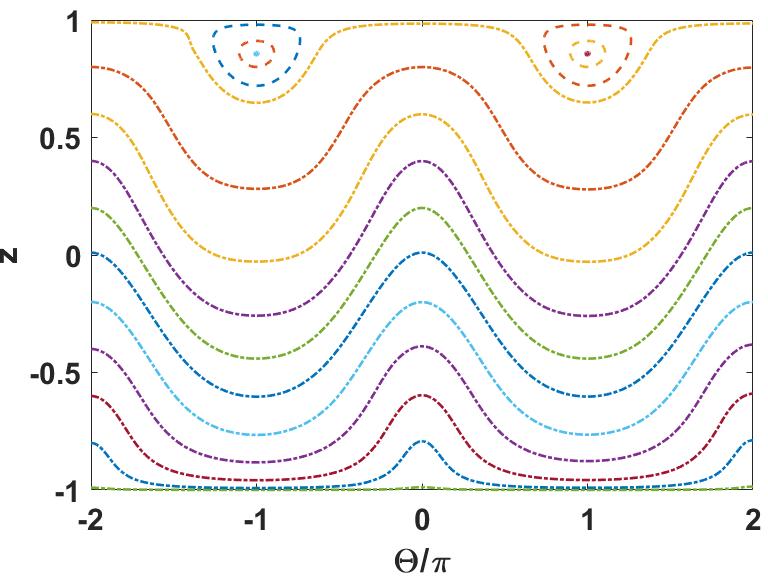} \\ c)}
\end{minipage}
\caption{Phase plane $z-\Theta$ at %(a) $\Delta = 0.5$, $\Omega = 0.05\pi$; (b) $\Delta = 0.5$, $\Omega = \pi$; (c) $\Delta = 0.5$, $\Omega = 1.5\pi$; (d) $\Delta = 0.75$, $\Omega = 0.05\pi$; (e) $\Delta = 0.75$, $\Omega = \pi$; (f) $\Delta = 0.75$, $\Omega = 1.5\pi$.}
$\Delta = 0.75$ and for (a) $\Omega = 0.05\pi$; (b) $\Omega = \pi$; (c) $\Omega = 1.5\pi$.}
\label{FIG:PD_Omega}
\end{figure*}

The dynamical behavior of the coupled soliton system can be generalized in terms of a phase portrait of two dynamical variables $z$ and $\Theta$, as shown in Figs.~\ref{FIG:PD_Delta} and~\ref{FIG:PD_Omega}. 

In Fig.~\ref{FIG:PD_Delta} we represent $z-\Theta$ phase-plane for $\Omega=0$ and for different (increasing) values of distance $\Delta$. We distinguish three different dynamic regimes. Solid curves correspond to oscillation regime when $z(\tau)$ and $\Theta(\tau)$ are some periodic functions of normalized time, see e.g.~\eqref{lin_osc} and red curve in Fig.~\ref{FIG:CR}. The dashed curves in Fig.~\ref{FIG:PD_Delta} indicate the self-trapping regime, when $z(\tau)$ is periodic and the sign of $z$ does not change, c.f.~\eqref{self-trap} and blue curve in Fig.~\ref{FIG:CR}. Physically, this is the macroscopic quantum self-trapping (MQST) regime characterized by a nonzero average population imbalance, when the most of particles are "trapped" within one of the solitons. At the same time, the behavior of phase $\Theta(\tau)$ may be quite complicated but periodic in time. On the other hand, for the running-phase regime depicted by dashed-dotted curves $\Theta(\tau)$ grows infinitely, see green curve in Fig.~\ref{FIG:PD_Delta}(b). Due to the symmetry that takes place at $\Omega=0$, the running-phase can be achieved only with the MQST regime, see Fig.~\ref{FIG:PD_Delta}.

As seen from Fig.~\ref{FIG:PD_Delta}, central area of nonlinear Rabi-like oscillations between the ground and first excited macroscopic states happen for a relatively small inter-soliton distance $\Delta$ and are inherent to zero-phase oscillations, see Fig.~\ref{FIG:PD_Delta}(a). As we discussed before, at $\Delta=\Delta_c\approx0.5867$ this area splits into two regions characterized by the MQST regimes, Fig.~\ref{FIG:PD_Delta}(b). This splitting occurs due to the bifurcation of population imbalance, cf. black curve in Fig.~\ref{FIG:D(z)}. These regions are moving away from each other with growing $\Delta$, see Fig.~\ref{FIG:PD_Delta}(c-f). It is worth noticing the bifurcation effect and occurrence of MQST states at zero-phase regime for coupled solitons in Fig.~\ref{FIG:Solitons} disappear for the condensates described by Gaussian states, cf.~\cite{Ostrovskaya2000,Elyutin2001}. 

The phase trajectories inherent to $\pi$-phase region $\frac{\pi}{2}<\Theta<\frac{3\pi}{2}$ stay weakly perturbed until the second critical value $\Delta\approx2$, when the MQST regime in Fig.~\ref{FIG:PD_Delta}(d) changes to Rabi-like oscillations (Fig.~\ref{FIG:PD_Delta}(e)) and then approaches the running-phase at $\Delta\approx6$, see Fig.~\ref{FIG:PD_Delta}(f). 

At large enough $\Delta$ the particle tunneling vanishes and the zero-phase MQST domains arise in the vicinity of population imbalance $z=\pm1$, Fig.~\ref{FIG:PD_Delta}(f). The phase dynamics corresponds to the running-phase regime with $z = \const$, see Fig.~\ref{FIG:PD_Delta}(f) and~\eqref{large_Delta}.

For non-zero $\Omega$, the phase portrait becomes asymmetric, Fig.~\ref{FIG:PD_Omega}. To elucidate the role of $\Omega$ we study the soliton interaction for a given inter-soliton distance $\Delta=0.75>\Delta_c$ that corresponds to the one after the bifurcation. As seen from Fig.~\ref{FIG:PD_Omega}(a), the phase portrait does not change significantly for small $\Omega$, cf. Fig.~\ref{FIG:PD_Delta}(b).

One of the SS solutions for zero and $\pi$- phase regimes disappears with increasing $\Omega$; the running-phase regime establishes, Fig.~\ref{FIG:PD_Omega}(b). Further increasing of $\Omega$ leads to vanishing of the SS solution for zero-phase, Fig.~\ref{FIG:PD_Omega}(c). 

Thus, phase portraits in Figs.~\ref{FIG:PD_Delta},~\ref{FIG:PD_Omega} provide the existence of degenerate SSs for coupled solitons by varying inter-soliton distance $\Delta$ and $\Omega$. Such solutions, as we show below, may be exploited for the macroscopic superposition soliton states formation in the quantum approach. 

%The MQST regime occurs with increasing $\Omega$, see Fig.~\ref{FIG:PD_Omega}(b). For large enough $\Omega$, the running-phase state is observed, Fig.~\ref{FIG:PD_Omega}(c). In this case the region $\frac{\pi}{2}<\Theta<\frac{3\pi}{2}$ turns out to be more sensitive to $\Omega$ changing than $-\frac{\pi}{2}<\Theta<\frac{\pi}{2}$, see Fig.~\ref{FIG:PD_Omega}(a,b). It is important, that due to the symmetry breaking at $\Omega\neq0$, all the self-trapping regimes are not degenerated and the running-phase regime possesses $z(\tau)$ with the large amplitude, see Fig.~\ref{FIG:PD_Omega}(c). In this limit the phase portrait in Fig.~\ref{FIG:PD_Omega}(c) approaches one described by Gaussian-shape condensates~\cite{555}. \textbf{Figs.~\ref{FIG:PD_Omega}(d-f) demonstrate that for $\Delta>0.5867$ the phase portraits change slightly except the one with $\Omega=0.05\pi$}

\section{Quantum metrology with two-soliton states.}% measurements.}

\subsection{Phase estimation with macroscopic qubit states.}

Suppose that some quantum system is prepared in state $|\psi\rangle$, which carries information about some parameter $\Gamma$ that we would like to estimate; in this work we are interested in fundamental bound for POVM measurements and consider pure states of the quantum system.% in which the system can be prepared. 

In quantum metrology the sensitivity of some parameter $\Gamma$ estimation is described by the error propagation formula given as (cf.~\cite{Toth}) 
\begin{equation}\label{EP-formula}
 \sigma_\Gamma = \frac{\sqrt{\left\langle\psi|(\Delta\hat{\Pi})^2|\psi\right\rangle}}{ \left|\frac{\partial\left\langle\psi|\hat{\Pi}|\psi\right\rangle}{\partial\Gamma}\right|},
\end{equation}
where $\left\langle\psi|(\Delta\hat{\Pi})^2|\psi\right\rangle = {\left\langle\psi|\hat{\Pi}^2 |\psi\right\rangle - \left\langle\psi|\hat{\Pi}|\psi\right\rangle}^2$ is the variance of fluctuations of some operator $\hat{\Pi}$ that corresponds to the measurement procedure. Typically, such procedures are based on appropriate interferometric schemes and use quantum superpositions, which contain required information about estimated parameter $\Gamma$. In the case of SC-states, which presume macroscopic (non-orthogonal in general) states, the measurement procedure requires some specification. In particular, we assume that quantum system may be prepared in the state $|\psi\rangle$ that we represent as 
\begin{equation}\label{M-qubit1}
 |\psi\rangle=\frac{1}{\sqrt{2}}(|\pi_0\rangle+e^{- i\phi_N}|\pi_1\rangle).
\end{equation}
In~\eqref{M-qubit1} $\phi_N$ is a relative (estimated) phase between states $|\pi_0\rangle$ and $|\pi_1\rangle$, which are defined as 
\begin{subequations}\label{M-qubit}
\begin{eqnarray}
 |\pi_0\rangle=c_1|\Phi_1\rangle- c_2|\Phi_2\rangle;\\
 |\pi_1\rangle=c_2|\Phi_1\rangle- c_1|\Phi_2\rangle,
\end{eqnarray}
\end{subequations}
%In~\eqref{M-qubit} the $c_1$ and $c_2$ are real coefficients obeying condition $c_0^2+c_1^2=1$. 
$|\Phi_1\rangle$ and $|\Phi_2\rangle$ are two macroscopic states representing two "halves" of the SC-states. In particular, operators $\hat{\Pi}_i= |\pi_i\rangle\langle \pi_i|$ realize a projection onto the superposition of states $|\Phi_{1,2}\rangle$, which generally are not orthogonal to each other obeying the condition 
\begin{equation}\label{Normalization}
\left\langle \Phi_1\Big|\Phi_2\right\rangle =\eta. 
\end{equation}
Simultaneously, we require the states in Eqs.\eqref{M-qubit} to fulfill the normalization condition 
\begin{equation}\label{Norm2}
 \left\langle \pi_i\Big|\pi_j\right\rangle =\delta_{ij}, i,j=0,1. 
\end{equation}

Now, we are able to determine the coefficients $c_{1,2}$, which fulfill to Eqs.\eqref{Normalization},~\eqref{Norm2} and look like 
\begin{equation}\label{Coef}
c_{1,2}=\sqrt{\frac{1\pm\sqrt{1-\eta^2}}{2(1-\eta^2)}}. 
\end{equation}
In Eqs.\eqref{Normalization},~\eqref{Coef} parameter $\eta$ defines the distinguishability of the states $|\Phi_{1,2}\rangle$. The case of $\eta=1$ in~\eqref{Normalization},~\eqref{Coef} corresponds to completely indistinguishable states $|\Phi_{1,2}\rangle$. In this case one can assume that $|\Phi_{1}\rangle$ and $|\Phi_{2}\rangle$ represent the same state. 

On the other hand, the situation with $\eta=0$ characterizes in~\eqref{Normalization},~\eqref{Coef} completely orthogonal states $|\Phi_{1,2}\rangle$ that become possible if $|\Phi_{1,2}\rangle$ approaches two-mode Fock states. In other words, this is a limit of the $N00N$-state for which coupled solitons are examined. 

Then, we define a complete set of operators $\hat{\Sigma}_j$, $j=1,2,3$ (cf.~\cite{Gerry2007})
\begin{subequations}\label{Sigma1}
\begin{eqnarray}
\hat{\Sigma}_0&=&\left|\pi_1\right\rangle\left\langle\pi_1\right| + \left|\pi_0\right\rangle\left\langle\pi_0\right|, \\
\hat{\Sigma}_1&=&\left|\pi_1\right\rangle\left\langle\pi_1\right| - \left|\pi_0\right\rangle\left\langle\pi_0\right|, \\
\hat{\Sigma}_2&=&\left|\pi_0\right\rangle\left\langle\pi_1\right| + \left|\pi_1\right\rangle\left\langle\pi_0\right|, \\
\hat{\Sigma}_3&=&i( \left|\pi_0\right\rangle\left\langle\pi_1\right| - \left|\pi_1\right\rangle\left\langle\pi_0\right|), 
\end{eqnarray}
\end{subequations}
which obey the SU(2) algebra commutation relation. 
 
The meaning of sigma-operators is evident from their definitions~\eqref{Sigma1}. Due to properties~\eqref{Norm2} the states $|\pi_{i}\rangle$ are suitable candidates for macroscopic qubit states, which we can define by mapping $|\pi_{0}\rangle\rightarrow |\mathbf{0}\rangle$ and $|\pi_{1}\rangle\rightarrow |\mathbf{1}\rangle$, respectively, cf.~\cite{Izumi2018, Cochrane1999}. In this form we can use them for POVM measurements defined with operators~\cite{Nielsen2010} 
 \begin{subequations}\label{POVM}
\begin{eqnarray}
 E_1&\equiv&\frac{\sqrt{2}}{1+\sqrt{2}}|\mathbf{1}\rangle\left\langle\mathbf{1}\right|=\frac{1}{\sqrt{2}(1+\sqrt{2})}(\hat{\Sigma}_0+\hat{\Sigma}_1);\\
 E_2&\equiv&\frac{1}{\sqrt{2}(1+\sqrt{2})}(|\mathbf{0}\rangle-|\mathbf{1}\rangle)(\langle\mathbf{0}| + \langle\mathbf{1}|)=-\frac{1}{\sqrt{2}(1+\sqrt{2})}(\hat{\Sigma}_1+i\hat{\Sigma}_3); \\
 E_3&\equiv&I-E_1-E_2.
\end{eqnarray}
\end{subequations}
Importantly, current quantum (photonic) technologies permit POVM tomography, cf.~\cite{Lundeen2009}.
 
Average values of sigma-operators in~\eqref{Sigma1} may be obtained by means of~\eqref{M-qubit1},~\eqref{Norm2} and look like 
\begin{subequations}\label{temp11}
\begin{eqnarray}
 \left\langle\hat{\Sigma}_1\right\rangle&=&0,\\
 \left\langle\hat{\Sigma}_2\right\rangle&=&\cos[\phi_N],\\
 \left\langle\hat{\Sigma}_3\right\rangle&=&\sin[\phi_N].
\end{eqnarray}
\end{subequations}
From~\eqref{temp11} it follows that only $\left\langle\hat{\Sigma}_{2,3}\right\rangle$ contain the information about desired phase $\phi_N$.

To estimate the sensitivity of phase measurement it is possible to assume that $\phi_N=N\Gamma$ and use Eq.~\eqref{EP-formula} with measured operator $\hat{\Pi}\equiv \hat{\Sigma_2}$.
Taking into account $\left\langle\hat{\Sigma}_2^2\right\rangle = 1$ for the variance of fluctuations $\left\langle(\Delta\hat{\Sigma}_2)^2\right\rangle$ we obtain 
%\begin{equation}
 % \left\langle\hat{\Sigma}^2\right\rangle = 1,
%\end{equation}
\begin{equation}\label{var1}
 \left\langle(\Delta\hat{\Sigma}_2)^2\right\rangle = \sin^2[N\Gamma].
\end{equation}

Finally, from Eqs.~\eqref{EP-formula},~\eqref{var1} for the phase error propagation we obtain 
\begin{equation}\label{n00n_scale}
\sigma_\Gamma = \frac{1}{N},
\end{equation}
that clearly corresponds to HL of arbitrary ($N$-linearly dependent) phase estimation and explores the sigma-operator measurement procedure. Notice this procedure can be mapped onto the parity measurement, cf.~\cite{Tsarev2018,Gerry2007}.

%in the framework of POVM approach for and linearly depends on $N$. 

\subsection{Soliton SC-qubit states.}

The phase estimation procedure described above enables to use two-soliton quantum states~\eqref{GS} for frequency quantum metrology purposes. It is instructive to represent 
%which possess creation of various macroscopic superposition states which use degeneracy of population imbalance $z$. 
soliton wave functions~\eqref{anz_0} in the form 
\begin{subequations}\label{ansatz_z1}
\begin{eqnarray}
 \Psi_1&=&\frac{\sqrt{uN}}{4}(1-z)\sech\left[\left(1-z\right)\left(\frac{uN}{4}x-\Delta\right)\right]e^{-i\frac{\theta}{2}};\\
 \Psi_2&=&\frac{\sqrt{uN}}{4}(1+z)\sech\left[\left(1+z\right)\left(\frac{uN}{4}x+\Delta\right)\right]e^{i\frac{\theta}{2}}.
\end{eqnarray}
\end{subequations}
Degenerate SS solutions of population imbalance obtained before (see e.g.~\eqref{n00n_sol} and Fig.~\ref{FIG:D(z)}) enable to prepare various superposition soliton states for quantum metrology purposes. 
%Below we examine below two fundamental precision bounds for frequency and inter-soliton distance measurements with coupled solitons. 
%For linear phase-shift $\varphi$ measurement and estimation one can use degenerate zero or $\pi$-phase SS solutions $z=\pm z_0$ of Eq.~\eqref{eqs}. 
In particular, for 
%$\Delta\approx0$ and 
$\Theta=0$ from~\eqref{GS} we obtain 
%e.g.~\eqref{SC_z}) we deal here with SC states possessing some finite (non-zero) particle numbers in both of solitons. For the 
\begin{subequations}\label{HF_SC}
\begin{eqnarray}
 \left|\Phi_1\right\rangle&=&\frac{1}{\sqrt{N!}}\left[\int_{-\infty}^\infty\left(\Psi_1^{(+)}\hat{a}_1^\dag + \Psi_2^{(+)}\hat{a}_2^\dag\right)dx\right]^N\left|0\right\rangle,\\
 \left|\Phi_2\right\rangle&=&\frac{1}{\sqrt{N!}}\left[\int_{-\infty}^\infty\left(\Psi_1^{(-)}\hat{a}_1^\dag + \Psi_2^{(-)}\hat{a}_2^\dag\right)dx\right]^N\left|0\right\rangle,
\end{eqnarray}
\end{subequations}
for two "halves" of the SC-state, where %(upper sign in~\eqref{HF_SC}) and for $\Theta=\pi$ (lower sign in~\eqref{HF_SC}), respectively. In~\eqref{HF_SC} we defined 
%\textcolor{blue}{
\begin{subequations}\label{ansatz_z}
\begin{eqnarray}
 %\Psi_1^{(\pm)}&=&\frac{\sqrt{uN}}{4}(1\mp z_0)\sech\left[\left(1\mp z_0\right)\left(\frac{uN}{4}x - \Delta\right)\right];\\
 %\Psi_2^{(\pm)}&=&\frac{\sqrt{uN}}{4}(1\pm z_0)\sech\left[\left(1\pm z_0\right)\left(\frac{uN}{4}x + \Delta\right)\right],%;\\
 \Psi_1^{(\pm)}&=&\frac{\sqrt{uN}}{4}(1-z_{\pm})\sech\left[\left(1-z_{\pm}\right)\left(\frac{uN}{4}x - \Delta\right)\right];\\
 %\Psi_1^{(-)}&=&\frac{\sqrt{uN}}{4}(1-z_-)\sech\left[\left(1-z_-\right)\left(\frac{uN}{4}x - \Delta\right)\right];\\
 \Psi_2^{({\pm})}&=&\frac{\sqrt{uN}}{4}(1+z_{\pm})\sech\left[\left(1+z_{\pm}\right)\left(\frac{uN}{4}x + \Delta\right)\right].
 %\Psi_2^{(-)}&=&\frac{\sqrt{uN}}{4}(1+z_-)\sech\left[\left(1+z_-\right)\left(\frac{uN}{4}x + \Delta\right)\right].
\end{eqnarray}
\end{subequations}

In Eqs.~\eqref{ansatz_z} $z_{+}$ and $z_{-}$ are two SS solutions corresponding to upper and lower branches in Fig.~\ref{FIG:D(z)}, respectively. In~\eqref{ansatz_z} we omit the common unimportant term $e^{-iN(\theta/2 + \beta_1\tau)}$. In particular, for $\Omega\approx0$, we have $z_{\pm}\rightarrow \pm z_0$. 
%\begin{equation}
% \Psi_\pm&=&\frac{\sqrt{uN}}{4}(1\pm z_0)\sech\left[\left(1\pm z_0\right)\frac{uN}{4}x\right].
%\end{equation}

%Noticing, different sign in Eqs.\eqref{HF_SC} has no influence on a $\varphi$-measurement procedure. 
%Thus we treat this state as the same SC-state. 

The scalar product for state~\eqref{HF_SC} is 
\begin{equation}\label{eta}
\eta\equiv\left\langle \Phi_1\Big|\Phi_2\right\rangle = \left[\int_{-\infty}^\infty\left(\Psi_1^{(+)}\Psi_1^{(-)} + \Psi_2^{(+)}\Psi_2^{(-)}\right)dx\right]^N\equiv \epsilon^N,
 %\eta\equiv\left\langle \Phi_1\Big|\Phi_2\right\rangle = \left[2\int_{-\infty}^\infty\Psi_+\Psi_-dx\right]^N\equiv \epsilon^N.
\end{equation}
where $\epsilon$ characterizes solitons wave functions overlapping. 
Assuming non-zero and positive $\Omega$ for $\epsilon$ one can obtain 
\begin{eqnarray}\label{eps}
\epsilon&=&\frac{1}{2}\Bigg(\left(1-\frac{z_++z_-}{2}\right)\left(1-z_+\right)\left(1-z_-\right)\left(1-0.21\left[\frac{z_+-z_-}{2-(z_++z_-)}\right]^2\right)\nonumber\\
&+&\left(1+\frac{z_++z_-}{2}\right)\left(1+z_+\right)\left(1+z_-\right)\left(1-0.21\left[\frac{z_+-z_-}{2+(z_++z_-)}\right]^2\right)\Bigg).
\end{eqnarray}

\begin{figure}[h!]
\center{\includegraphics[width=0.5\linewidth]{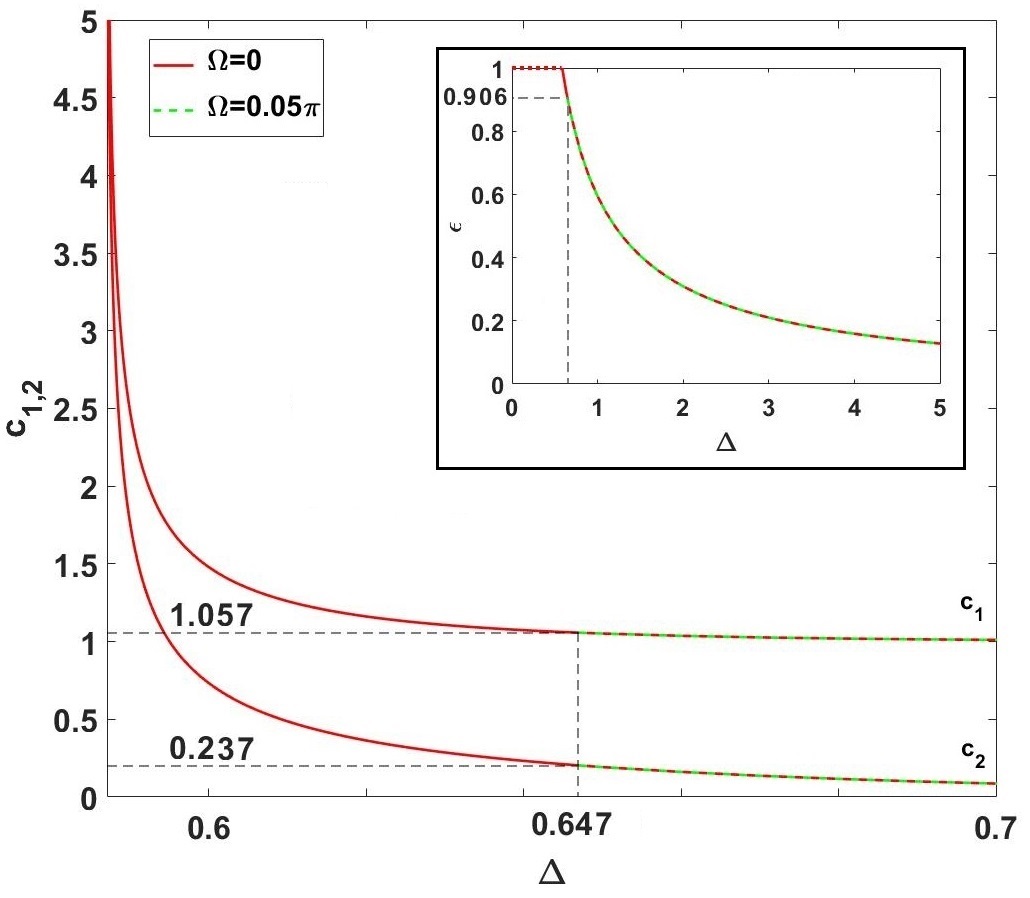}}
\caption{Coefficients $c_{1,2}$ versus normalized inter-soliton distance $\Delta$ for \textbf{$N=10$} and different $\Omega$. The inset demonstrates behavior of $\epsilon$ for different $\Delta$.}
\label{FIG:C_Delta}
\end{figure}

In Fig.~\ref{FIG:C_Delta} we establish the principal features of coefficients~\eqref{Coef} and parameter $\epsilon$ (see the inset in Fig.~\ref{FIG:C_Delta}) as functions of $\Delta$. The value of $\Omega$ plays a significant role in the distinguishability problem for states $|\Phi_{1}\rangle$ and $|\Phi_{2}\rangle$. In particular, at $\Omega=0$, in the bifurcation point $\Delta=\Delta_c=0.5867$ we have $\epsilon=1$ that implies indistinguishable states $|\Phi_{1}\rangle$ and $|\Phi_{2}\rangle$, see red curve in the inset of Fig.~\ref{FIG:C_Delta}. The coefficients $c_{1,2}\rightarrow\infty$ in this limit. 

However, even for the small (positive) $\Omega$ it follows from~\eqref{eps} that $\epsilon\neq1$ for any $\Delta>\Delta_c$, and states $|\Phi_{1}\rangle$, $|\Phi_{2}\rangle$ are always distinguishable. In particular, it follows from zero-phase solution~\eqref{2ODE_2} that $z_{\pm} = \pm\left(1.2 - \frac{18\Delta_-}{\omega^2}\right)\sqrt{\Delta_-} - \Omega\frac{f(\Delta_-)}{\omega^2}$ and $|z_{+}|\neq|z_{-}|$. This situation is displayed by green curves in Fig.~\ref{FIG:C_Delta}; the SS solutions possess $c_1=1.057$, $c_2=0.203$ for $c_{1,2}$ that correspond to $\Delta\approx0.647$, $\epsilon\approx0.9056$ for $\Omega=0.05\pi$. 

From Fig.~\ref{FIG:C_Delta} it is evident that the coefficients $c_{1,2}$ rapidly approach (due to the factor $N$) the levels $c_{1}=1$, $c_{2}=0$ (completely distinguishable macroscopic SC soliton states), when $\Delta$ increases. In this limit, as seen from Fig.~\ref{FIG:D(z)}, $z_{\pm}$ approaches $\pm z_0$, and from~\eqref{eps} we obtain
%small (and negative) shift $\delta_z$, $|\delta_z|<<z_0$ assuming that characterizes deviation of population imbalance from $z_0$, i.e. from equilibrium population imbalance $\Omega\approx0$, cf. Fig.~\ref{FIG:D(z)}. 
\begin{equation}\label{Size}
%\epsilon \approx(1-z_0^2)(1-0.21z_0^2) -\frac{\delta_z^2}{2}(1-0.21z_0^2).
\epsilon \approx(1-z_0^2)(1-0.21z_0^2).
\end{equation}
Practically, in this limit red and green curves coincide in Fig.~\ref{FIG:C_Delta}. 

%epsilon = \left(1 - z_0^2\right)\int_{0}^\infty\frac{dx}{\cosh^2\left[x\right] + \sinh^2\left[z_0x\right]}\approx(1-z_0^2)(1-0.21z_0^2) -\frac{\delta_z^2}{2}(1-0.21z_0^2).
%\end{equation}

Thus, we can exploit states~\eqref{HF_SC} for metrological measurement purposes for arbitrary phase $\phi_N$ estimation that we describe in Sec.~5.1. The phase $\phi_N$ may be created after soliton SC-state formation by means of some additional soliton interaction or collisions. 

\subsection{Frequency measurement, $\Gamma\equiv\Omega$.}

Now we represent a particularly important case of measurement of angular frequency $\Omega$ that characterizes energy spacing between the ground and first excited macroscopic states. The SS solutions~\eqref{n00n_sol}, which correspond to the maximal population imbalance $z^2=1$, allow to prepare the maximally path-entangled superposition state, a.k.a. $N00N$-state. As seen from~\eqref{N00N_b}, the solution with $z=1$ exist, when $-2(\pi-1)\leq\Omega/\Lambda\leq2(\pi+1)$. Similarly, the domain of solution $z=-1$ is $-2(\pi+1)\leq\Omega/\Lambda\leq2(\pi-1)$. To achieve the superposition $N00N$-state formation we require both solutions to exist simultaneously. This restricts the domain of allowed $\Omega$ as $-2(\pi-1)\leq\Omega/\Lambda\leq2(\pi-1)$.

Substituting $z=\pm1$ into~\eqref{ansatz_z1} we obtain 
\begin{subequations}\label{n00n_Psi}
\begin{eqnarray}
 \Psi_1&=&\frac{\sqrt{uN}}{2}\sech\left[\left(\frac{uN}{2}x-2\Delta\right)\right]e^{-i\frac{\theta}{2}},\\
 \Psi_2&=&\frac{\sqrt{uN}}{2}\sech\left[\left(\frac{uN}{2}x+2\Delta\right)\right]e^{i\frac{\theta}{2}},
\end{eqnarray}
\end{subequations}
which are relevant to the $N00N$-state's two "halves" defined as 
%if $z=1$ all the $N$ particles populate the second soliton, possessing $\Psi_1=0$ and $\Psi_2=\Psi e^{\frac{i}{2}\theta}$, where
%\begin{equation}\label{n00n_Psi}
% \Psi=\frac{\sqrt{uN}}{2}\sech\left[\left(\frac{uN}{2}x-2\Delta\right)\right].
%\end{equation}
%On the other hand, from~\eqref{ansatz_z1} for $z=-1$ we obtain $\Psi_2=0$ and $\Psi_1=\Psi e^{-\frac{i}{2}\theta}$. Substituting $z=\pm1$ and~\eqref{n00n_Psi} into~\eqref{GS} we obtain two "halves" of the $N00N$-state
\begin{subequations}\label{HF_N00N}
\begin{eqnarray}
\left|0N\right\rangle&=&\frac{1}{\sqrt{N!}}\left[\int_{-\infty}^\infty\Psi_1\hat{a}_1^\dag dx\right]^N\left|0\right\rangle,\\
\left|N0\right\rangle&=&\frac{1}{\sqrt{N!}}\left[\int_{-\infty}^\infty\Psi_2\hat{a}_2^\dag dx\right]^N\left|0\right\rangle.
\end{eqnarray}
\end{subequations}
Considering the superposition of states~\eqref{HF_N00N} and omitting unimportant common phase $e^{iN\left(0.5\theta^{(+)}-\beta_2t\right)}$ we arrive to (cf.~\eqref{M-qubit1})
%\begin{equation}\label{N00N-state_0}
 % \left|N00N\right\rangle = \frac{1}{\sqrt{2}}\left(e^{iN\left(0.5\theta^{(+)}-\beta_2t\right)}\left|N0\right\rangle + e^{iN\left(-0.5\theta^{(-)}-\beta_1t\right)}\left|0N\right\rangle\right),
%\end{equation}
%where $\theta^{(\pm)}$ is solitons' phase-shift at $\sign[z]=\pm1$. 
%As seen from~\eqref{n00n_sol} the effective phase-shift $\Theta$ depends on the sign of $z$, but $\Omega=\beta_2-\beta_1$ does not. Thus the sign of $z$ must be taken into account in $\theta$ which is done in~\eqref{N00N-state_0}. 
 %we finally write down the $N00N$-state as
\begin{equation}\label{N00N-state}
 \left|N00N\right\rangle = \frac{1}{\sqrt{2}}\left(\left|N0\right\rangle + e^{- iN\Theta'}\left|0N\right\rangle\right),
\end{equation}
that represents the $N00N$-state of coupled BEC solitons for our problem. In~\eqref{N00N-state} 
\begin{equation}\label{Theta-eff}
 \Theta' = \frac{\Theta^{(+)} + \Theta^{(-)}}{2} = \frac{1}{2}\left(\arccos\left[\frac{2\Lambda - \Omega}{2\pi\Lambda}\right] + \arccos\left[\frac{2\Lambda + \Omega}{2\pi\Lambda}\right]\right)
\end{equation}
is the phase shift that contains $\Omega$-parameter required for estimation. 
%dependent that is relevant to appropriate measurement procedure. 
%Noticing that the $N00N$-state~\eqref{N00N-state} exist within the domain $-2\Lambda(\pi-1)\leq\Omega\leq2\Lambda(\pi-1)$, see~\eqref{Theta-eff}.
%The phase-shift~\eqref{Theta-eff} depends on $\Omega$, so this parameter can be measured via the state~\eqref{N00N-state}. 
%For the sake of simplicity we assume that inter-soliton distance is small enough, $\Delta<<1$. 
Comparing Eq.~\eqref{Theta-eff} with~\eqref{M-qubit1} we can conclude that the $N00N$-state's "halves" $\left|N0\right\rangle$ and $\left|0N\right\rangle$ in~\eqref{Theta-eff} may be associated with states $|\pi_0\rangle$ and $|\pi_1\rangle$, respectively. To estimate the sensitivity of $\Omega$ measurement we use~\eqref{EP-formula} with measured operator $\hat{\Pi}\equiv \hat{\Sigma}$ defined as (cf.\eqref{Sigma1}(c))
\begin{equation}\label{Sigma}
 \hat{\Sigma} = \left|N0\right\rangle\left\langle0N\right| + \left|0N\right\rangle\left\langle N0\right|.
\end{equation}
Since the states~\eqref{HF_N00N} are orthogonal, the mean value of~\eqref{Sigma} is
\begin{equation}\label{temp1}
 \left\langle\hat{\Sigma}\right\rangle = \cos[N\Theta'].
\end{equation}

\begin{figure*}[h!]
\center{\includegraphics[width=0.5\linewidth]{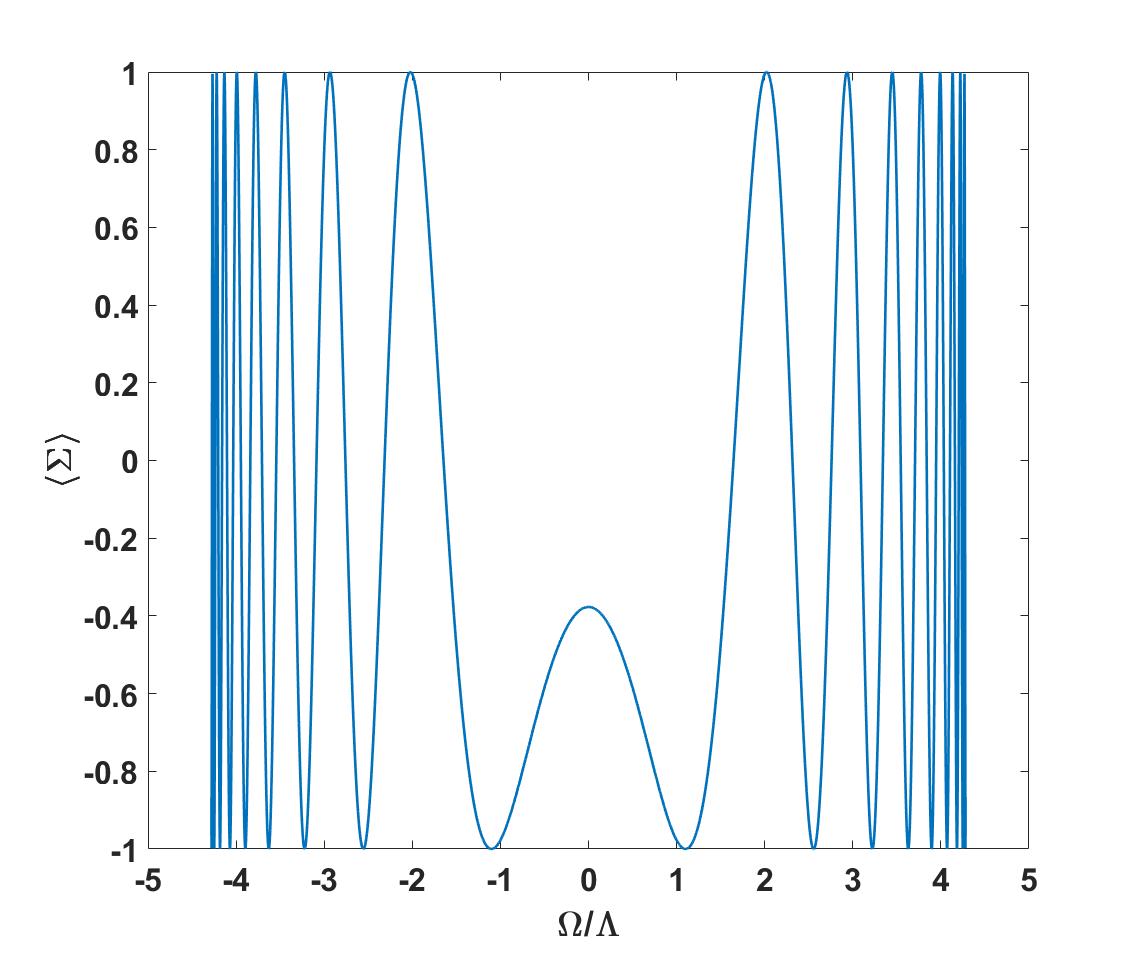}}
\caption{Mean value $\left\langle\hat{\Sigma}\right\rangle$ vs. normalized angular frequency $\Omega/\Lambda$ for %(a)$N=10$; (b) $N=50$; (c) $N=100$; (d) 
$N=200$.}
\label{FIG:Sigma}
\end{figure*}

Fig.~\ref{FIG:Sigma} demonstrates $\left\langle\hat{\Sigma}\right\rangle$ as a function of $\Omega/\Lambda$. Notice, the interference pattern in Fig.~\ref{FIG:Sigma} exhibits essentially nonlinear behavior for measured $\left\langle\hat{\Sigma}\right\rangle$.

The variance of fluctuations $\left\langle(\Delta\hat{\Sigma})^2\right\rangle$ for the measured sigma-operator reads as 
%\begin{equation}
 % \left\langle\hat{\Sigma}^2\right\rangle = 1,
%\end{equation}
\begin{equation}\label{var}
 \left\langle(\Delta\hat{\Sigma})^2\right\rangle = \sin^2[N\Theta'].
\end{equation}

Now, by using~\eqref{EP-formula},~\eqref{var} we can easily find the propagation error for frequency $\Omega$ estimation as
%According to estimation theory~\cite{}, the error of measurement $\Omega$ via the operator~\eqref{Sigma} can be estimated as
%\begin{equation}
 % \sigma_\Omega = \sqrt{\left\langle\left(\Delta\Omega\right)^2\right\rangle} = \sqrt{\left\langle\left(\Delta\hat{\Sigma}\right)^2\right\rangle/\left|\frac{\partial\left\langle\hat{\Sigma}\right\rangle}{\partial\Omega}\right|^2}=\frac{1}{N}\left|\frac{\partial\Theta'}{\partial\Omega}\right|^{-1},
%\end{equation}
%which gives the scaling of the sensitivity 
\begin{equation}\label{n00n_scale2}
 \sigma_\Omega = \frac{2\Lambda}{N}\left|\frac{\sqrt{4\pi^2-(2+\Omega/\Lambda)^2}\sqrt{4\pi^2-(2-\Omega/\Lambda)^2}}{\sqrt{4\pi^2-(2+\Omega/\Lambda)^2} - \sqrt{4\pi^2-(2-\Omega/\Lambda)^2}}\right|,
\end{equation}

Equation~\eqref{n00n_scale2} is non-applicable for $\Omega=0$ since the denominator in~\eqref{n00n_scale2} turns to zero. We choose the optimal estimation area for the frequency $\Omega$, where 
%The scaling~\eqref{n00n_scale2} is plotted in Fig.~\ref{FIG:scaling}. As seen, 
the best sensitivity is reached, in the vicinity of the domain border at $\Omega/\Lambda\rightarrow2(\pi-1)$. In this limit~\eqref{n00n_scale2} approaches 
\begin{equation}\label{n00n_scale_2}
 %\sigma_\Omega \simeq \frac{4\Lambda}{N}\frac{\sqrt{\pi(\pi-1)}\sqrt{2(\pi-1) - \Omega}}{1.65 - \sqrt{\pi}\sqrt{4.28 - \Omega)}}.
 \sigma_\Omega \simeq \frac{10\Lambda}{N}\frac{1.65\sqrt{4.28 - \Omega/\Lambda}}{1.65 - \sqrt{4.28 - \Omega/\Lambda}}.
\end{equation}

Equation~\eqref{n00n_scale_2} exhibits one of the important results of this work: for a given $\Lambda$ that characterizes atomic condensate peculiarities Eq.~\eqref{n00n_scale_2} demonstrates Heisenberg scaling for frequency measurement sensitivity, cf.~\eqref{n00n_scale}.

%\begin{figure*}[ht]
%\center{\includegraphics[width=0.5\linewidth]{8_sigma_Omega.jpg}}
%\caption{$\Omega$ measurement scaling: direct calculation~\eqref{n00n_scale} (blue solid line) and the Taylor expansion~\eqref{n00n_scale_2} (red dashed line).}
%\label{FIG:scaling}
%\end{figure*}

\section*{Conclusion}

In summary, we have considered the problem of two-soliton formation for 1D BECs trapped effectively in a double-well potential. 
%We consider bright matter wave soliton formation in the framework of coupled-mode theory for BECs. The soltions characterized by condensate macroscopic ground and first excited states superposition. 
%The interaction between solitons appear due to the nonlinear Josephson effect that in our case leads to non-trivial particle tunneling behavior between the solitons. In particular, such a tunneling may be characterized by some functionals $J$ and $I$ which depend on normalized inter-soliton distance $\Delta$ and angular frequency $\Omega$ that describes energy spacing between ground and first excited macroscopic states of the condensate. 
These soliton Josephson junctions analytical solutions and corresponding phase portraits exhibit the occurrence of novel macroscopic quantum selft-trapping (MQST) phases in contrast to the condensates with only Gaussian wave functions. With these soliton states, we have also explored the formation of the Schr{\"o}dinger-cat (SC) state in the framework of the Hartree approximation. In particular, we have analyzed the distinguishabiltiy problem for binary (non-orthogonal) macroscopic states. Compared to the results known in the literature, see e.g.~\cite{Tsarev2018}, finite frequency spacing $\Omega$ leads to the distinguishable macroscopic states for condensate solitons. This circumstance may be important for the experimental design of the SC-states. 

The important part of this work is devoted to the applicability of predicted states for quantum metrology. By utilizing the macroscopic qubits problem with interacting BEC solitons, one can apply the sigma-operators to elucidate the measurement and subsequent estimation of arbitrary phase, that linearly depends on the particle number, up to the HL. It is worth mentioning the sigma-operators relate to the POVM detection tomography procedure. On the other hand, the phase estimation procedure for the phase-dependent sigma-operator can be realized by means of the parity measurement technique that produces the same accuracy for phase estimation. We have shown that in the limit of soliton state solution with the population imbalance $|z|=\pm1$ the coupled soliton system admits the maximally path-entangled $N00N$-state formation. The feasibility of frequency $\Omega$ estimation at the Heisenberg level is also demonstrated. 

In this paper we have not examined the loses and decoherence effects for the quantum soliton system depicted in Fig.~\ref{FIG:Solitons}. Recently in~\cite{Tsarev2020} we examined this problem for quantum solitons possessing simple Josephson coupling. From the experimental point of view, the recent BEC soliton experiments with lithium condensates demonstrated, that the collisions may be recognized as one of the most detrimental effects~\cite{Nguyen2014}. In particular, as we established in~\cite{Tsarev2020}, the three-body and one-body losses may be unimportant at the time scales of few tens of milliseconds that relevant to experimental conditions in~\cite{Nguyen2014}. Moreover, the purely quantum analysis of the problem demonstrated that the superposition of Fock states, occurring at some specific parameters of the system, behaves robust to few particle losses. The Detailed analysis of this problem for interaction of solitons depicted in Fig.~\ref{FIG:Solitons} we will publish elsewhere. 

\section*{Funding}

This work was partially supported by the Ministry of Science and Technology of Taiwan under Grant Nos. MOST 108-2923-M-007-001-MY3 and 109-2112-M-007-019-MY3 and financially supported by the Grant of RFBR, No 19-52-52012 MHT{\_}a.

\section*{Disclosures}

The authors declare no conflicts of interest.

\section*{Appendix: Approximation of functionals $I$ and $J$.}

\begin{figure*}[h!]
\begin{minipage}[h]{0.49\linewidth}
\center{\includegraphics[width=1\linewidth]{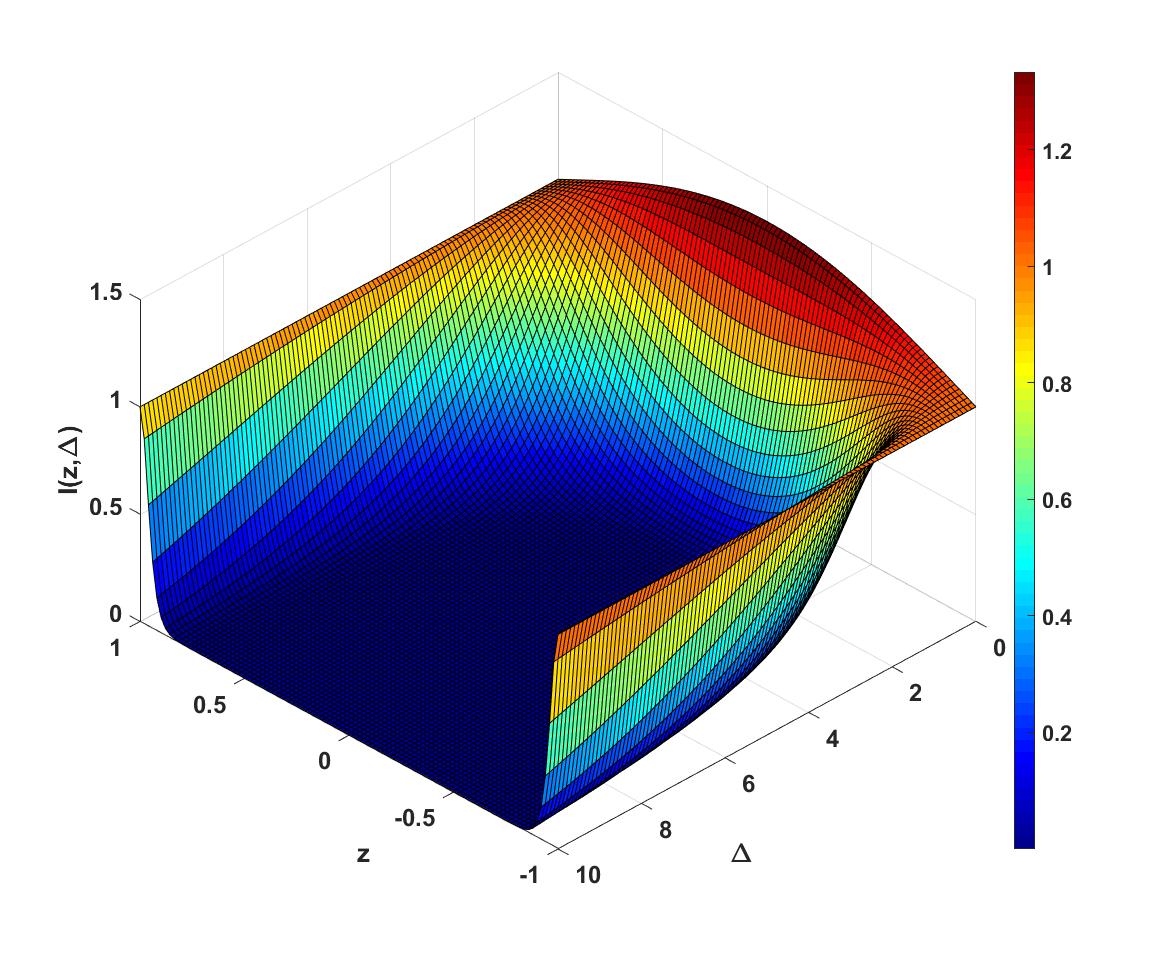} \\ a)}
\end{minipage}
\hfill
\begin{minipage}[h]{0.49\linewidth}
\center{\includegraphics[width=1\linewidth]{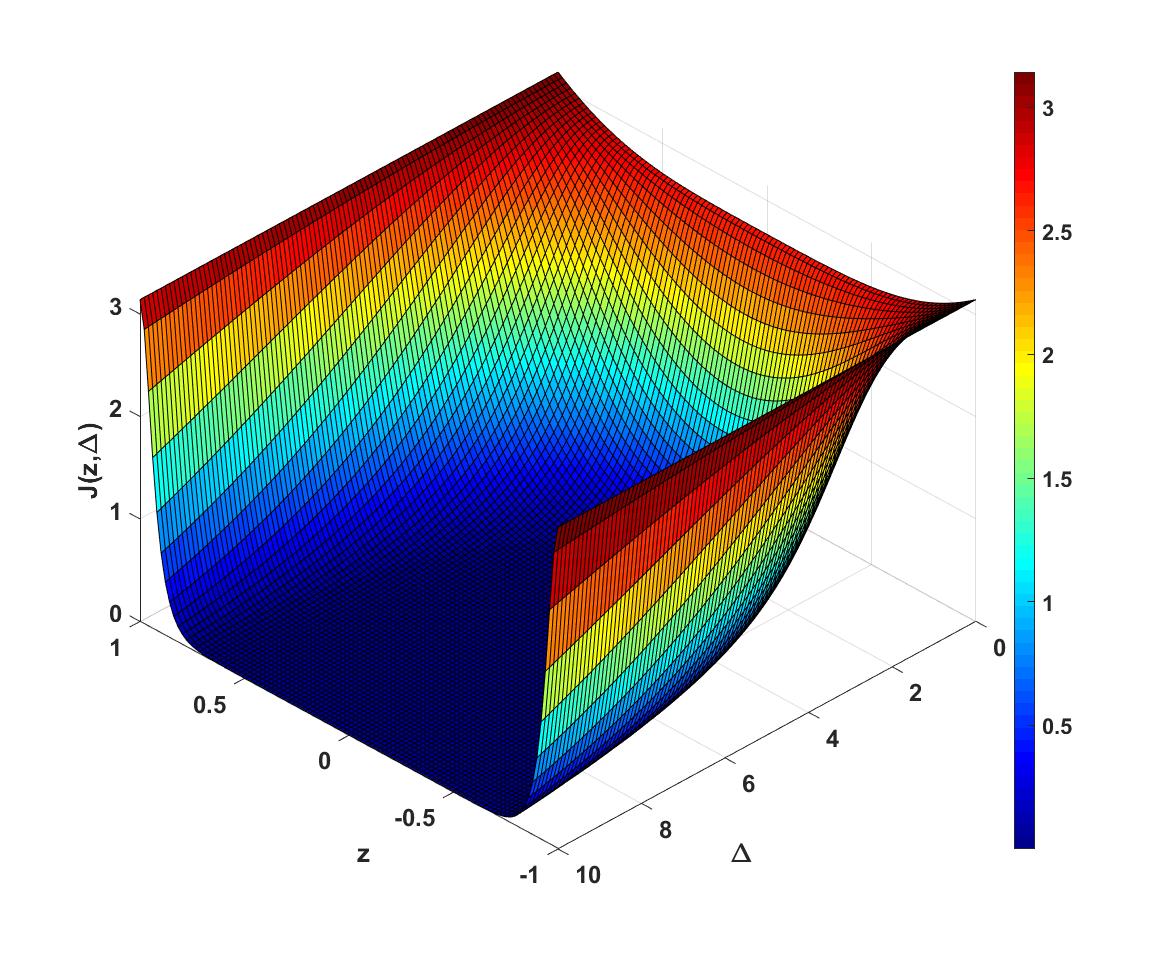} \\ b)}
\end{minipage}
\vfill
\begin{minipage}[h]{0.49\linewidth}
\center{\includegraphics[width=1\linewidth]{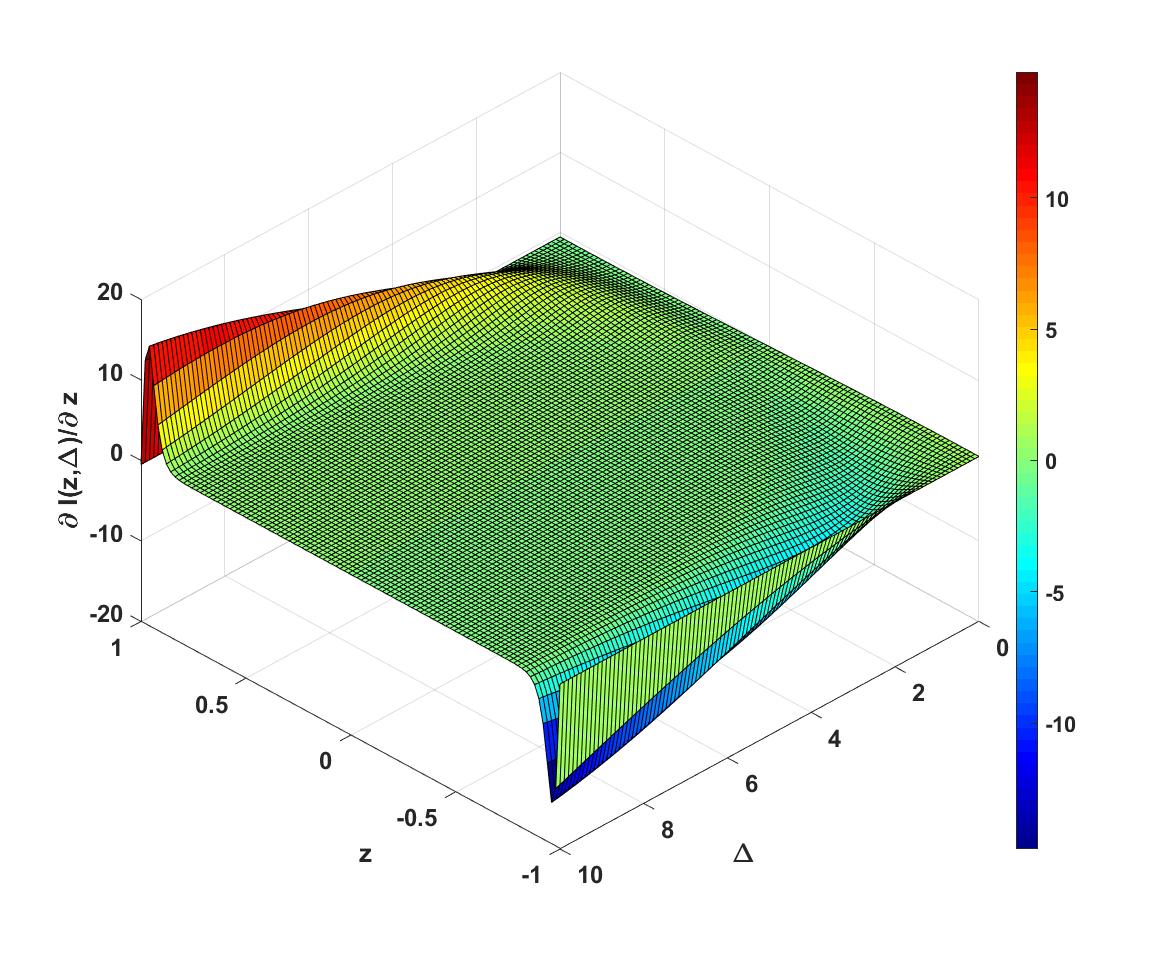} \\ c)}
\end{minipage}
\hfill
\begin{minipage}[h]{0.49\linewidth}
\center{\includegraphics[width=1\linewidth]{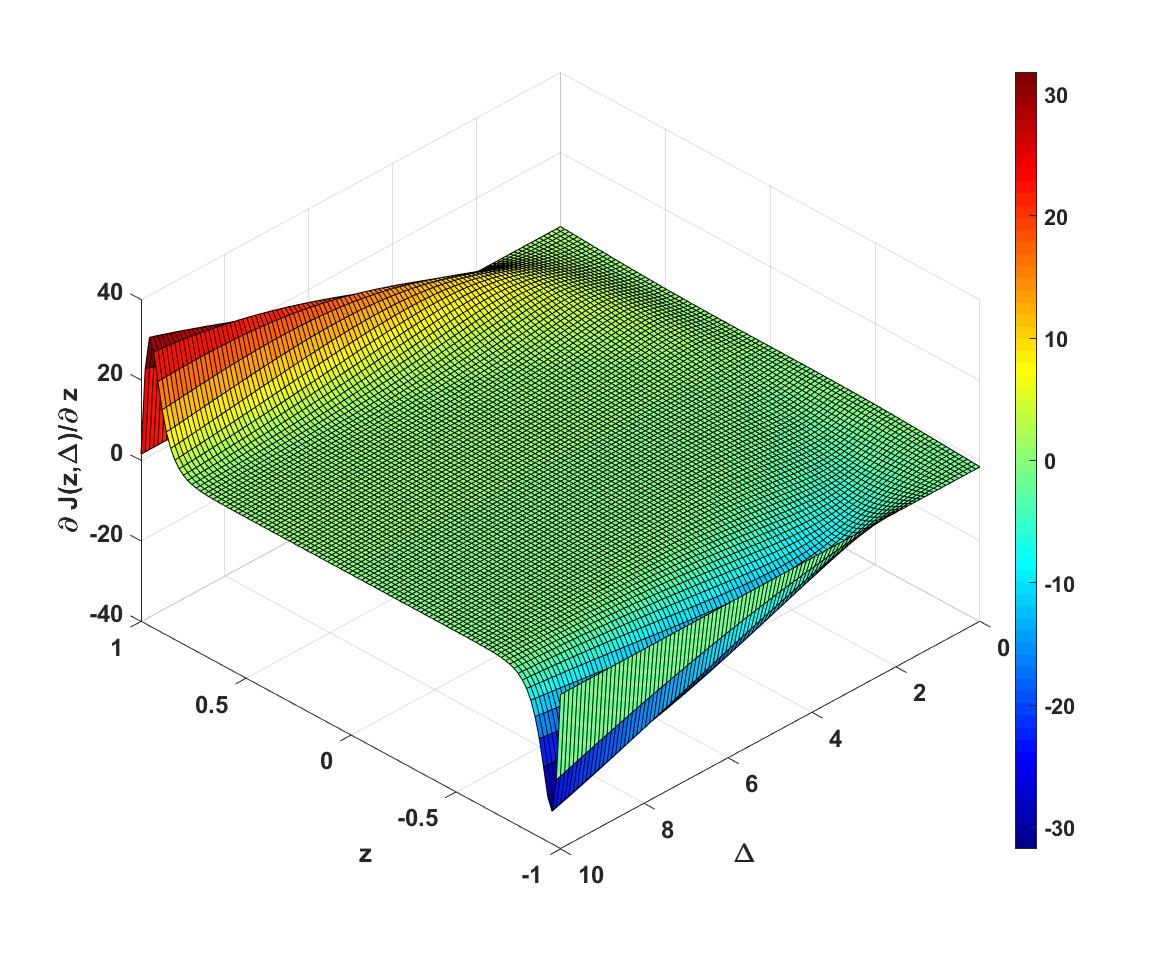} \\ d)}
\end{minipage}
\caption{(a-b) The $I,J$ functionals and (c-d) their derivatives with respect to $z$,$I',J'$, versus $z$ and $\Delta$.}
\label{FIG:functionals}
\end{figure*}

Solutions of Eqs.~\eqref{eqs} strictly depend on functionals $I$ and $J$ and their derivatives $I'\equiv dI/dz$ and $J'\equiv dJ/dz$ defined in~\eqref{funcs}. In Fig.~\ref{FIG:functionals} we represented them as two-dimension surfaces given in $z-\Delta$ plane. From Fig.~\ref{FIG:functionals} (a,b) it is seen that $I,J$ approaches zero for large $\Delta$ excluding edge domains where $|z|\simeq1$. The behavior of $I,J$, as it is follows from Fig.~\ref{FIG:functionals} (a,b), is not so evident for small $\Delta$ inherent to $0\leq\Delta<1.5$. Thus for numerical estimations we use the polynomial approximations of $I$ and $J$.

%We split this domain into two parts to create approximations for $I$ and $J$.

%In Sections II and III of this paper we use a polynomial approximation of the functionals~\eqref{funcs}. As seen from Fig.~\ref{FIG:functionals} it is enough to consider $0<\Delta\leq1.5$ to investigate the main dynamics features. 

In particular, within the domain $0\leq\Delta<0.6$, $I,J(z)$ can be effectively approximated by the forth-order polynomials as follows
\begin{subequations}\label{A1}
\begin{eqnarray}
I(z,\Delta)&\approx&a_I(\Delta)z^4+b_I(\Delta)z^2+c_I(\Delta);\\
J(z,\Delta)&\approx&a_J(\Delta)z^4+b_J(\Delta)z^2+c_J(\Delta),
\end{eqnarray}
\end{subequations}
where the coefficients are the polynomials themselves:
\begin{subequations}
\begin{eqnarray}
a_I&=&-\Delta^2-0.52\Delta+0.1;\\
b_I&=&2\Delta^2+0.76\Delta-0.42;\\
c_I&=&-1.16\Delta^2-0.24\Delta+1.33;\\
a_J&=&-2\Delta^2-0.72\Delta+0.4;\\
b_J&=&3.9\Delta^2+1.03\Delta+0.07;\\
c_J&=&-1.9\Delta^2-0.32\Delta+2.7.
\end{eqnarray}
\end{subequations}

At $0.6<\Delta\leq1.5$ in~\eqref{A1} the sixth-order polynomial approximation is required, that is 
\begin{subequations}\label{A2}
\begin{eqnarray}
I(z,\Delta)&\approx&a_I(\Delta)z^6 + b_I(\Delta)z^4 + c_I(\Delta)z^2 + d_I(\Delta);\\
J(z,\Delta)&\approx&a_J(\Delta)z^6 + b_J(\Delta)z^4 + c_J(\Delta)z^2 + d_J(\Delta),
\end{eqnarray}
\end{subequations}
where
\begin{subequations}
\begin{eqnarray}
a_I&=&0.31\Delta^2 - 2.57\Delta + 1.43;\\
b_I&=&0.9\Delta^2 + 1.24\Delta - 1.6;\\
c_I&=&-1.9\Delta^2 + 3.5\Delta - 0.67;\\
d_I&=&0.69\Delta^2 - 2.21\Delta + 1.85;\\
a_J&=&-1.5\Delta^2 - 0.13\Delta + 0.89;\\
b_J&=&4.62\Delta^2 - 4.78\Delta + 0.15;\\
c_J&=&-4\Delta^2 + 8.4\Delta - 1.45;\\
d_J&=&0.94\Delta^2 - 3.52\Delta + 3.56.
\end{eqnarray}
\end{subequations}

An approximation that we use here provides error less than $4\%$ for any $-1\leq z\leq1$ and $\Delta$ in the mentioned domains.

%We use Eqs.~\eqref{A1},~\eqref{A2} to elucidate SS solutions at $\Theta=0,\pi$ and $\Delta \neq0$. 

%In particular, for $\Omega=0$ and $\Theta=0$ the Eq.~\eqref{eqs_b} reduces to:
%\begin{equation}\label{A3}
% 0 = z\left(-12.5 + 14.94 z^2 - 7.98 z^4 + 1.2 z^6 + \Delta^2(24.56 - 60.56 z^2 + 48 z^4 - 12 z^6) + \Delta(6.42 - 20.68 z^2 + 20.52 z^4 - 6.24 z^6)\right)
%\end{equation}
%that valid in the domain $0<\Delta<0.6$, and to equation
%\begin{eqnarray}\label{A4}
% 0&=&z\Big(-21.13 + 15.94 z^2 + 40.08 z^4 - 60.64 z^6 + 21.45 z^8 + \Delta(47.6 - 100.54 z^2 + 13.95 z^4 + 77.6 z^6 - 38.55 z^8)\nonumber\\
% &+&\Delta^2(-19.72 + 66.82 z^2 - 67.23 z^4 + 15.36 z^6 + 4.65 z^8)\Big)
%\end{eqnarray}
%within $0.6\leq\Delta\leq1.5$ area of $\Delta$.
%Similarly, the conditions for $\Theta=\pi$ are 
%\begin{equation}\label{A5}
% 0 = z\left(-1.98 + 12.3 z^2 - 3.18 z^4 + 1.2 z^6 + \Delta^2(1.36 - 13.36 z^2 + 24 z^4 - 12 z^6) + \Delta(1.02 - 6.68 z^2 + 11.88 z^4 - 6.24 z^6)\right).
%\end{equation}
%at $0<\Delta<0.6$ and 
%\begin{eqnarray}\label{A6}
% 0&=&z\Big(-1.09 + 3.14 z^2 + 31.2 z^4 - 46.4 z^6 + 21.45 z^8 + \Delta(-0.08 + 4.9 z^2 - 41.85 z^4 + 75.52 z^6 - 38.55 z^8)\nonumber\\
% &+&\Delta^2(0.04 - 2.14 z^2 + 6.21 z^4 - 8.64 z^6 + 4.65 z^8)\Big)
%\end{eqnarray}

%The Eqs.~\eqref{A3}-\eqref{A6} has a solution $z=0$ for any $\Delta$. In addition, Eqs.~\eqref{A3},~\eqref{A4} provide SS solutions plotted in Fig.~\ref{FIG:D(z)}. At the same time Eqs.~\eqref{A5},~\eqref{A6} provide degenerate self-trapped states $z=\pm z_0$ suitable for the SC-state formation, see Fig.~\ref{FIG:PD_Delta}. 

\end{document}